\documentclass{aa}

\usepackage{lscape}

\bibpunct{(}{)}{;}{a}{}{,} 
\newcommand{\teff}{\ensuremath{T_{\rm eff}}}
\def\subsun{\mbox{$_{\normalsize\odot}$}}
\newcommand{\logg}{\ensuremath{\log{g}}}
\newcommand{\zfe}{[Fe/H]}

\usepackage{hyperref}
\hypersetup{colorlinks=true}
\hypersetup{citecolor=blue}
\hypersetup{linkcolor=blue}
\hypersetup{urlcolor=blue}

\usepackage{graphicx}
\usepackage[varg]{txfonts}
\defcitealias{2015A&A...577A..42B}{BHAC15}
\defcitealias{pre013}{Pass18}
\begin{document}

\title{The CARMENES search for exoplanets around M~dwarfs\thanks{Table B.1 is only available in electronic form
at the CDS via anonymous ftp to cdsarc.u-strasbg.fr (130.79.128.5) or via \url{http://cdsweb.u-strasbg.fr/cgi-bin/qcat?J/A+A/}}}

\subtitle{Different roads to radii and masses of the target stars}

\titlerunning{The CARMENES exoplanet search around M~dwarfs: Different roads to radii and masses of the targets}

\author{A. Schweitzer
\inst{\ref{hs}} \and V.~M.~Passegger\inst{\ref{hs}}
\and C.~Cifuentes\inst{\ref{cab},}\inst{\ref{ucm}}
\and V.~J.~S.~B\'ejar\inst{\ref{iac}}
\and M.~Cort\'es-Contreras\inst{\ref{cab}}
\and J.~A.~Caballero\inst{\ref{cab}}
\and C.~del~Burgo\inst{\ref{inaoe}}
\and S.~Czesla\inst{\ref{hs}}
\and M.~K{\"u}rster\inst{\ref{mpia}}
\and D.~Montes\inst{\ref{ucm}}
\and M.~R.~Zapatero~Osorio\inst{\ref{cab2}}
\and I.~Ribas\inst{\ref{ice}, \ref{ieec}}
\and A.~Reiners\inst{\ref{iag}}
\and A.~Quirrenbach\inst{\ref{lsw}}
\and P.~J.~Amado\inst{\ref{iaa}}
\and J.~Aceituno\inst{\ref{iaa}, \ref{caha}}
\and G.~Anglada-Escud\'e\inst{\ref{iaa}, \ref{qmul}}
\and F.~F.~Bauer\inst{\ref{iaa}}
\and S.~Dreizler\inst{\ref{iag}}
\and S.~V.~Jeffers\inst{\ref{iag}}
\and E.~W.~Guenther\inst{\ref{tls}}
\and T.~Henning\inst{\ref{mpia}}
\and A.~Kaminski\inst{\ref{lsw}}
\and M.~Lafarga\inst{\ref{ice}, \ref{ieec}}
\and E.~Marfil\inst{\ref{ucm}}
\and J.~C.~Morales\inst{\ref{ice}, \ref{ieec}}
\and J.~H.~M.~M.~Schmitt\inst{\ref{hs}}
\and W.~Seifert\inst{\ref{lsw}}
\and E.~Solano\inst{\ref{cab}}
\and H.~M.~Tabernero\inst{\ref{cab2}, \ref{ucm}}
\and M.~Zechmeister\inst{\ref{iag}}
}

\institute{Hamburger Sternwarte, Gojenbergsweg 112, D-21029 Hamburg, Germany, \email{aschweitzer@hs.uni-hamburg.de} \label{hs}
\and Departamento de Astrof\'isica, Centro de Astrobiolog\'ia (CSIC-INTA), ESAC Campus, Camino Bajo del Castillo s/n, E-28691 Villanueva de la Ca\~nada, Madrid, Spain \label{cab}
\and Departamento de F\'{i}sica de la Tierra y Astrof\'{i}sica and UPARCOS-UCM (Unidad de F\'{i}sica de Part\'{i}culas y del Cosmos de la UCM), Facultad de Ciencias F\'{i}sicas, Universidad Complutense de Madrid, E-28040, Madrid, Spain  \label{ucm}
\and Instituto de Astrof\'{\i}sica de Canarias, V\'ia L\'actea s/n, E-38205 La Laguna, Tenerife, Spain, and Departamento de Astrof\'{\i}sica, Universidad de La Laguna, E-38206 La Laguna, Tenerife, Spain \label{iac}
\and Instituto Nacional de Astrof\'{\i}sica, \'Optica y Electr\'onica, Luis Enrique Erro 1, Sta. Ma. Tonantzintla, Puebla, Mexico \label{inaoe}
\and Max-Planck-Institut f\"ur Astronomie, K\"onigstuhl 17, D-69117 Heidelberg, Germany \label{mpia}
\and Centro de Astrobiolog\'ia (CSIC-INTA), Carretera de Ajalvir km 4, 28850 Torrej{\'o}n de Ardoz, Madrid, Spain \label{cab2}
\and Institut de Ci\`encies de l'Espai (CSIC-IEEC), Can Magrans s/n, Campus UAB, E-08193 Bellaterra, Barcelona, Spain \label{ice}
\and Institut d'Estudis Espacials de Catalunya (IEEC), C/ Gran Capit{\`a} 2-4, 08034 Barcelona, Spain \label{ieec}
\and Institut f\"ur Astrophysik, Georg-August-Universit\"at, Friedrich-Hund-Platz 1, D-37077 G\"ottingen, Germany \label{iag}
\and Landessternwarte, Zentrum f\"ur Astronomie der Universt\"at Heidelberg, K\"onigstuhl 12, D-69117 Heidelberg, Germany\label{lsw}
\and Instituto de Astrof\'isica de Andaluc\'ia (IAA-CSIC), Glorieta de la Astronom\'ia s/n, E-18008 Granada, Spain \label{iaa}
\and Centro Astron\'omico Hispano-Alem\'an (CSIC-MPG), Observatorio Astron\'omico  de  Calar  Alto,  Sierra  de  los  Filabres, E-04550  G\'ergal, Almer\'ia, Spain\label{caha}
\and School of Physics and Astronomy, Queen Mary, University of London, 327 Mile End Road, London, E1 4NS, United Kingdom\label{qmul}
\and Th\"uringer Landessternwarte Tautenburg, Sternwarte 5, D-07778 Tautenburg, Germany\label{tls}
}

\date{Received 21 December 2018 / Accepted 25 March}

\abstract{}{
We determine the radii and masses of 293 nearby, bright M~dwarfs of the CARMENES survey.
This is the first time that such a large and homogeneous high-resolution ($R>80,000$) spectroscopic survey
has been used to derive these fundamental stellar parameters.
}{
We derived the radii using Stefan-Boltzmann's law. We obtained the required effective temperatures \teff\ from
a spectral analysis and we obtained the required luminosities $L$ from integrated broadband photometry
together with the {\em Gaia} DR2 parallaxes.
The mass was then determined using a mass-radius relation that
we derived from eclipsing binaries known in the literature.
We compared this method with three other methods:
(1) We calculated the mass from the radius and the surface gravity \logg, which was
    obtained from the same spectral analysis as \teff.
(2) We used a widely used infrared mass-magnitude relation.
(3) We used a Bayesian approach to infer stellar parameters from the comparison of the absolute magnitudes and colors of our targets with evolutionary models.
}{
Between spectral types M0\,V and M7\,V our radii cover the range $0.1\,R\subsun<R<0.6\,R\subsun$ with
an error of 2--3\% and our masses cover $0.09\,{\mathcal M}\subsun<{\mathcal M}<0.6\,{\mathcal M}\subsun$ with
an error of 3--5\%.
We find good agreement between the masses determined with these different methods for most of our targets. Only the
masses of very young objects show discrepancies. This can be well explained
with the assumptions that we used for our methods.
}{}


   \maketitle

\section{Introduction}

The mass of a star is one of its most important properties.
When trying to understand a star by itself and when isolating it
from its environment, it is fundamental to
stellar physics that the mass is the
most influential parameter that determines almost exclusively all other properties throughout
a star's life.
Moreover, should the star have a companion, the masses of both components determine the
gravitational potential and their orbits.
In particular, if the companion is a planet, the gravitational potential is dominated
by the stellar mass, which is crucial to know
if we want to measure the minimum mass, ${\mathcal M}_{\rm P} \sin{i}$, of the planet.

The latter is the situation that the CARMENES 
search faces in its quest to find Earth-mass planets.
CARMENES (Calar Alto high-Resolution search for M~dwarfs with Exo-earths
with Near-infrared and optical \'Echelle Spectrographs)
is a double channel spectrograph built to find Earth-mass planets around M~dwarfs using
the radial-velocity method \citep{2014SPIE.9147E..1FQ}.
The visual channel of this spectrograph has a resolving power of $R\approx95\,000$ and covers the wavelength range from 0.52\,$\mu$m to 0.96\,$\mu$m.
Its near infrared channel  has a resolving power of $R\approx80\,000$ and covers the wavelength range from 0.96\,$\mu$m to 1.71\,$\mu$m.
This spectrograph is operated at the 3.5\,m telescope on Calar Alto, Spain, and has been taking data since January 2016 \citep{2016SPIE.9908E..12Q,2018SPIE10702E..0WQ,pre021}.
So far CARMENES\ has been used in the discovery of several new planets \citep{pre023,pre025,pre022,pre030,pre036}. All of these discoveries estimated the mass of the host star with the same procedure that we present in this work, even though \citet{pre023}, \citet{pre025}, and \citet{pre022}
used slightly older data than those used in this work.
However, an in-depth investigation of the validity of this method is still missing.

A widely used method to determine masses of M~dwarfs directly is the
observation of detached double-lined eclipsing binary systems 
\citep[see, e.g.,][for reviews]{1991A&ARv...3...91A,2010A&ARv..18...67T,2013AN....334....4T}.
Accurately measuring 
the radial-velocity amplitudes of the components of such
systems allows us to derive their stellar masses independently.
   Another method to measure the masses of M~dwarfs dynamically are  observations of astrometric binaries that are also
visual or spectroscopic binaries \citep[see, e.g.,][and both their references]{2001IAUS..200..539Q,2016AJ....152..141B},
for which it is possible to reconstruct the orbits of the components.

For measuring stellar radii, eclipsing binaries can be used as well.
The radii of their components and the inclination of the system can be determined from
the eclipsing light curve \citep[e.g.,][]{2002ApJ...567.1140T,2003A&A...398..239R,2009A&A...502..253C,2009ApJ...691.1400M}.
For single stars, interferometry is used to
directly measure the angular diameter
\citep[e.g.,][]{2009MNRAS.394.1925V,2012ApJ...746..101B,2012ApJ...757..112B,2016A&A...586A..94L}.
Together with the distance, the physical
radius can be inferred from the same dataset. However, it is difficult to apply this method to
low-mass stars because they are small and faint, which requires high
sensitivity and resolution to achieve acceptable accuracy. Some interferometric
radius measurements of M~dwarfs with uncertainties of 1--5\% have been performed
by \citet{2003A&A...397L...5S}, \citet{2006ApJ...644..475B}, \citet{2012ApJ...757..112B},
and \citet{2014MNRAS.438.2413V}.

On the other hand, because the fundamental techniques of measuring the mass and radius of isolated
M~dwarfs (or those with wide companions) are limited, well-calibrated empirical relationships are generally
employed. This is an adequate solution if the stars are assumed to behave like
statistical representatives of the sample on which these relations are based.
It has the advantage that only the luminosity or magnitude in certain bands has
to be measured to estimate the stellar mass. An early mass-luminosity relation
for M~dwarfs was provided by \citet{1993AJ....106..773H}.
Several other works
\citep[e.g.,][]{2000A&A...364..217D,2016AJ....152..141B,2018arXiv181106938M}
provided up-to-date mass-magnitude relations for
low-mass stars, calibrated for different parameter ranges. Additionally,
relations connecting colors or other measurable quantities with the mass or radius
that are not directly accessible have been established to determine missing parameters
of stars in question \citep[e.g.,][]{2015ApJ...804...64M}.
If there are, however, systematic differences between the samples
used to determine any such empirical relation and the stars
to be analyzed, then there is a bias in the results, which are based on such relations.
Most prominently, the level of stellar activity may cause such
a bias. For example, \citet{2010A&A...520A..91C}, \citet{2018MNRAS.476.3245J},
and \citet{2018AJ....155..225K}
reported inflated radii for young, magnetically active, or fast rotating stars.

Therefore, to obtain the mass  $\mathcal{M}$ of the CARMENES targets
we would traditionally use, for example, a ${\mathcal M}-M_K$ relation
\citep[e.g.,][]{2000A&A...364..217D} and an infrared magnitude.
This was actually done in the first CARMENES-based planet analysis \citep{pre015}
and in the survey overview of \citet{pre021}.
Our new approach, however, is to use such relations only when unavoidable.
Instead, we exploit as much information from the observations as possible
and use all the measurements, including our spectra and up-to-date photometry
and parallaxes, already obtained.
We also investigate the accuracy with which we can measure the radius and mass
of a star. This includes calculating individual error bars for every
analyzed object.
We finally confront our method with other methods (both new and traditional)
using the same homogeneous data set.

In Section~\ref{sec:sample} we present the sample used in this work. In Section~\ref{sec:method}
we describe our method.
In short, we combine several purely observational data sets (high-resolution spectroscopy, photometry, and trigonometric parallaxes)
with only one computational model (synthetic spectra based on a grid of model atmospheres)
and one fundamental relation (a mass-radius relation) to obtain luminosities, radii, and masses.
In Section~\ref{sec:compare} we compare the results from our method with
results obtained from other alternative methods. In Sections~\ref{sec:discussion} and \ref{sec:conclusion}
we finally discuss the results and draw conclusions for the future.

\section{The sample}
\label{sec:sample}
To date, the catalog of targets observed during guaranteed time observations (GTO)
of the CARMENES spectrograph includes a total of 341
bright, nearby M~dwarfs or M~dwarf systems, for which CARMENES based
results have been published, and we use 293 of these
for the reasons outlined below.

In this work we based our mass and radius determinations on the
spectral analysis of \citet[][see Section~\ref{sec:temperatures}]{2016A&A...587A..19P,pre013}.
We imposed the same requirements and only used those CARMENES spectra
with a signal-to-noise ratio (S/N) of S/N~$>$~75.
Furthermore, as in \citet[][hereafter Pass18]{pre013} we excluded targets with spectral line profiles that cannot
be reproduced by a purely photospheric model, i.e., stars
with signs of activity that significantly affect the
profiles of the lines used by \citetalias{pre013} for determining stellar
parameters; see \citet{2018A&A...614A.122T} and \citet{2018A&A...615A..14F}
for the impact of activity on CARMENES target stars.
Finally, as in \citetalias{pre013} we excluded fast rotators.
All these criteria finally yield 293 stars.

Previously, however, different CARMENES GTO samples have been studied.
\citet{pre021} presented the initial sample of 324 CARMENES targets
without any presumable companion closer than 5 arcsec, while
\citet{pre028} added nine double-lined spectroscopic binaries discovered
in the survey.
\citet{pre028} determined their minimum masses and radii using methods
suitable for spectroscopic binaries, but we do not consider these nine binaries
in this work because we are interested only in isolated M~dwarfs (or those with wide
companions).
One of the original 324 targets was recently determined
to be a close astrometric binary as well (to be discussed in a future publication) and we do not consider it in this work either.
There can still be some very long-period, low-amplitude, single-lined
spectroscopic binaries in our GTO sample that are yet to be discovered and did
not pass our initial filters \citep{2017A&A...597A..47C,2018A&A...614A..76J,pre028}.
Another eight objects did not have any or not sufficient
observations to reach enough S/N to
be tabulated by \citet{pre021}. Three of these\footnote{J15474-108 (GJ~3916), J18198-019 (Gl~710) and J20556-140S (Gl~810B)}
were included in the sample of \citet{pre019}, who studied the rotation periods
of CARMENES targets.
All of these eight objects except one\footnote{J18198-019 (Gl~710)} were included
in the study of \citet{pre034}, who investigated the activity of the CARMENES targets.
We, finally, included six\footnote{J04219+213 (K2-155), J08409-234 (Gl~317), J09143+526 (Gl~338A),
J15474-108 (GJ~3916), J18198-019 (Gl~710) and J20556-140S (Gl~810B)} of the eight new targets.

\begin{figure}
\centering
\includegraphics[width=1.0\hsize]{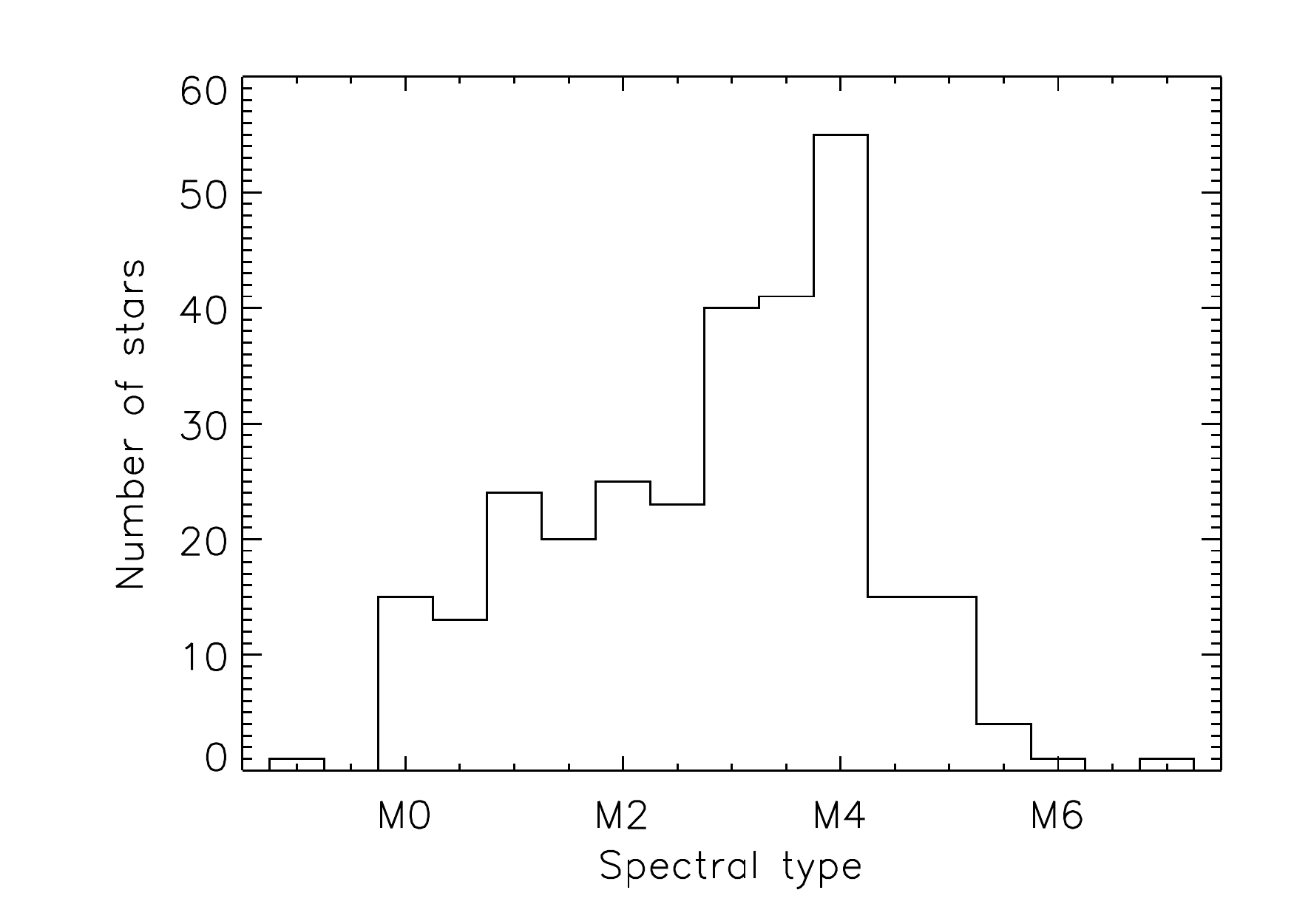}
   \caption{Histogram of the spectral types \citep[as determined by][and references therein]{2015A&A...577A.128A}
            of all 293 targets of our sample. The leftmost bin represents the only K7\,V star in our sample.
           }
      \label{fig:hist}
\end{figure}

We present our sample in Table~\ref{tab:results}.
For each of the 293 M~dwarfs, we provide for identification purposes
the CARMENES identifier \citep{2016csss.confE.148C},
the Gliese or Gliese \& Jahreiss name when available,
or the discovery name otherwise; we also give the luminosity-class V
spectral type \citep[][and references therein]{2015A&A...577A.128A}.
A histogram of the spectral types of our sample is shown in Fig.~\ref{fig:hist}.

\section{Our method}
\label{sec:method}

\begin{figure*}
\centering
\includegraphics[width=0.8\hsize]{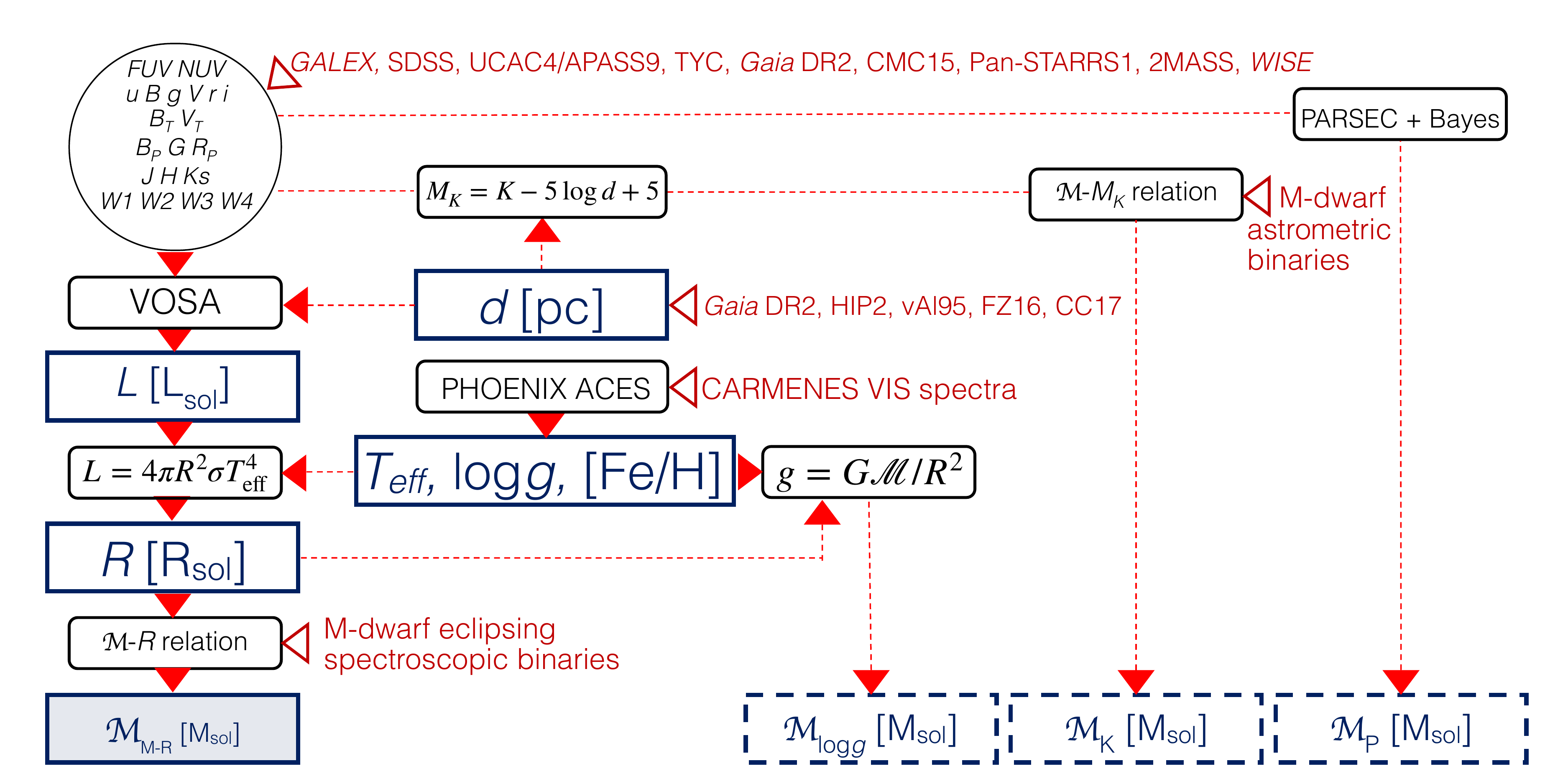}
   \caption{Flowchart of our different roads to masses.
           }
      \label{fig:methods}
\end{figure*}

Our method is a multi-step process.
In the first step in Section~\ref{sec:luminosities} we determine the luminosity
from distance and photometry. In the second step in Section~\ref{sec:temperatures}
we measure the effective temperature from the CARMENES spectra by fitting PHOENIX \citep{1999JCoAM.109...41H}
synthetic spectra. In the third step in Section~\ref{sec:radii} we use Stefan-Boltzmann's
law to obtain the radius.
To finally obtain the masses in Section~\ref{sec:mmr}, we
introduce an empirical mass-radius relation to calculate the mass
from the radius. These steps, along with those needed for the alternative methods,
are sketched in Fig.~\ref{fig:methods}.

\subsection{Distances and luminosities}
\label{sec:luminosities}

We determine the luminosities from
apparent magnitudes and distances.
For 285 of the 293 M~dwarfs, we retrieve parallactic distances from the {\em
Gaia} DR2 catalog \citep{2018arXiv180409365G}. For the remaining eight cases, which
are not part of {\em Gaia} DR2 and are listed in
Table~\ref{tab:results} together with a corresponding note,
we retrieve parallactic distances from the astrometric
analyses of
\citet[][two stars\footnote{J22020-194 (Gl~843) and J18224+620 (GJ~1227)}]{1995gcts.book.....V}, \citet[][three stars\footnote{J07274+052 (Gl~273), J11033+359 (Gl~411) and J11054+435 (Gl~412A)}]{2007A&A...474..653V},
and \citet[][two stars\footnote{J06574+740 (2M~J06572616+7405265) and J09133+688 (G~234-057)}]{2016AJ....151..160F},
and derive
spectrophotometric distances after applying the $J$-band absolute
magnitude-spectral type relation given by
\citet[][one star\footnote{J06396-210 (LP~780-032)}]{2017A&A...597A..47C}.

For every target star in the CARMENES input catalog Carmencita \citep{2016csss.confE.148C},
we collected apparent magnitudes in
broadband passbands from a number of all-sky and ultra-wide surveys
from the ultraviolet, through  optical and near-infrared, to the
mid-infrared.
The integration of the spectral energy distribution (SED) of each
CARMENES target allowed us to determine the bolometric luminosity $L$.
In our case, we performed the integration with the Virtual Observatory
SED Analyzer \citep[VOSA;][]{vosa}, the distances described above,
and the SEDs built with photometric data from
{\em GALEX} \citep[$FUV$ $NUV$; ][]{2007ApJS..173..682M},
SDSS \citep[$ugriz$; ][]{2000AJ....120.1579Y},
UCAC4 and APASS9 \citep[$BgVri$; ][]{2013AJ....145...44Z,2015AAS...22533616H},
Tycho-2 \citep[$B_T$ $V_T$; ][]{2000A&A...355L..27H},
{\em Gaia} DR2 \citep[$B_P$ $G$ $R_P$; ][]{2018arXiv180409365G},
CMC15 \citep[$r^\prime$;][]{2014AN....335..367M},
Pan-STARRS1 \citep[$grizy$; ][]{2012ApJ...750...99T},
2MASS \citep[$JHK_s$; ][]{2006AJ....131.1163S},
and
{\em AllWISE} \citep[$W1$ $W2$ $W3$ $W4$; ][]{2014yCat.2328....0C}.
The VOSA software uses the latest zero-point fluxes provided
by the Filter Profile Service \citep{2012ivoa.rept.1015R}, which can be found on their
website\footnote{\tt \url{http://svo2.cab.inta-csic.es/theory/fps/}}.
We used synthetic photometry based on the BT-Settl models \citep{2012RSPTA.370.2765A} for the ultraviolet flux
blueward of the Johnson $B$ band and for the mid- and far-infrared
redward of {\em WISE} $W4$.
At the considered effective temperatures, M~dwarfs
hardly emit any flux in these two wavelength regions.
Therefore, the approximation of these two contributions to the luminosity
are barely dependent on the models chosen as input to VOSA.
Moreover, by using synthetic photometry for the ultraviolet we do not
contaminate the photospheric flux with chromospheric flux from stellar
activity.
The resulting luminosities are listed in Table~\ref{tab:results}.
Typical relative errors of $L$ are between 1 and 2\%.
More details on the photometry compilation and luminosity derivation
including their errors
will be provided in a separate publication.

\subsection{Effective temperatures, surface gravities, and metallicities}
\label{sec:temperatures}
We measured the photospheric parameters effective temperature \teff, 
surface gravity \logg, and metallicity, denoted by the relative iron abundance \zfe\ 
by fitting PHOENIX-ACES synthetic spectra \citep{2013A&A...553A...6H}
to the CARMENES spectra of the visual (VIS) channel as described in \citet{2016A&A...587A..19P}
and \citetalias{pre013}.
In summary, this method has \teff\ and  \zfe\ as free parameters, but fixes \logg\ by
a \teff-\logg\ relation from theoretical 5~Gyr isochrones of \citet{1998A&A...337..403B}.
To analyze all 293 members of our sample described in Section~\ref{sec:sample},
we repeated the analysis of \citetalias{pre013} using the latest available CARMENES co-added \citep{pre018} VIS spectra
that fulfill the requirements of \citetalias{pre013} described above (cf. Section~\ref{sec:sample}). 

\citetalias{pre013} presented the stellar photospheric parameters for 235 stars
of our exoplanet survey.
These authors used CARMENES VIS data for 234 targets and CAFE (Calar Alto Fiber-fed Echelle spectrograph) data for one star.
We present in this work 59 new parameter sets
derived from CARMENES spectroscopy, and update the
parameters for 234 stars of \citetalias{pre013} (omitting the binary mentioned in Section~\ref{sec:sample}).
The updated and \citetalias{pre013}
results agree within their fixed error bars ($\Delta T_{\rm eff}$ =
51\,K, $\Delta \log{g}$~=~0.07\,dex, and $\Delta$[Fe/H]~=~0.16\,dex).
All results are listed in Table~\ref{tab:results}.

There exist, however, alternative model atmospheres and synthetic spectra
for M~dwarfs, most prominently the BT-Settl models \citep{2012RSPTA.370.2765A}. When they
are used in spectral analyses, the fit quality \citep[see, e.g.,][]{2012ApJ...748...93R}
is similar to the fit quality in \citetalias{pre013}.
However, the resulting photospheric parameters typically differ. Most recently, \citet{2018A&A...620A.180R}
determined effective temperatures for the CARMENES GTO targets listed in \citet{pre021} using the latest
BT-Settl models, which are about 100--200\,K smaller than ours.
Describing these differences is beyond the scope of this paper and they will be 
investigated in a future publication. However, in order to assess an error estimate
resulting from the use of one specific set of synthetic spectra we repeated
our analysis described above using the latest, publicly available
grid with varying metallicities, which are the 2011 BT-Settl models. We then found
effective temperatures that are about 50\,K (i.e., our $\Delta T_{\rm eff}$)
hotter for $\teff\lesssim$3700\,K and up to about 200\,K cooler for $\teff\gtrsim$3700\,K.
Owing to error propagation, this means that for
$\teff\lesssim$3700\,K the derived radii and masses (see below) that are based on PHOENIX-ACES models, and
radii and masses that are based on BT-Settl models would agree within our error bars.
However, for higher \teff\ values, there
would be a systematic offset between these derived quantities.
Similarly, using the significantly and systematically differing \teff\ results from  \citet{2018A&A...620A.180R}
would also introduce a systematic offset in radii and masses beyond our error bars.

\subsection{Radii}

\begin{table*}
\caption{Radii of 11 stars of our sample with interferometric diameters
measured by \citet{2012ApJ...757..112B}.}
\label{tab:rinterf}
\centering
\begin{tabular}{l l r@{$\pm$}l r@{$\pm$}l r@{$\pm$}l r@{$\pm$}l}
\hline\hline
Karmn     & Name   & \multicolumn{2}{c}{$R$}  &  \multicolumn{2}{c}{$R_{\rm interf}$} &  \multicolumn{2}{c}{$\Theta_{\rm LD}$\tablefootmark{B}}  & \multicolumn{2}{c}{$d$} \\           &        & \multicolumn{2}{c}{[$R\subsun$]}  &  \multicolumn{2}{c}{[$R\subsun$]} &  \multicolumn{2}{c}{[mas]}  & \multicolumn{2}{c}{[pc]} \\ \hline
 J00183+440 & Gl 15A & 0.393 & 0.011 & 0.385 & 0.002 & 1.005 & 0.005 & 3.5626 & 0.0005\tablefootmark{G}  \\
 J05314-036 & Gl 205 & 0.549 & 0.029 & 0.578 & 0.002 & 0.943 & 0.004 & 5.7003 & 0.0023\tablefootmark{G}  \\
 J09143+526 & Gl 338A & 0.578 & 0.020 & 0.568 & 0.010 & 0.834 & 0.014 & 6.3339 & 0.0016\tablefootmark{G}  \\
 J09144+526 & Gl 338B & 0.584 & 0.019 & 0.583 & 0.011 & 0.856 & 0.016 & 6.3335 & 0.0016\tablefootmark{G}  \\
 J11033+359 & Gl 411 & 0.359 & 0.016 & 0.392 & 0.004 & 1.432 & 0.013 & 2.5469 & 0.0043\tablefootmark{HIP}  \\
 J11054+435 & Gl 412A & 0.375 & 0.011 & 0.398 & 0.009 & 0.764 & 0.017 & 4.8480 & 0.0235\tablefootmark{HIP}  \\
 J11421+267 & Gl 436 & 0.427 & 0.013 & 0.437 & 0.014 & 0.417 & 0.013 & 9.7561 & 0.0086\tablefootmark{G}  \\
 J13457+148 & Gl 526 & 0.470 & 0.015 & 0.488 & 0.008 & 0.835 & 0.014 & 5.4354 & 0.0015\tablefootmark{G}  \\
 J15194-077 & Gl 581 & 0.308 & 0.010 & 0.302 & 0.009 & 0.446 & 0.014 & 6.2992 & 0.0020\tablefootmark{G}  \\
 J17578+046 & Gl 699 & 0.185 & 0.006 & 0.187 & 0.001 & 0.952 & 0.005 & 1.8267 & 0.0010\tablefootmark{G}  \\
 J22565+165 & Gl 880 & 0.527 & 0.015 & 0.549 & 0.003 & 0.744 & 0.004 & 6.8677 & 0.0019\tablefootmark{G}  \\
 \hline
\end{tabular}
\tablebib{
(B)~\citet{2012ApJ...757..112B};
(G)~\citet{2018arXiv180409365G};
(HIP)~\citet{2007A&A...474..653V} }
\end{table*}

\label{sec:radii}
\begin{figure}
\centering
\includegraphics[width=1.0\hsize]{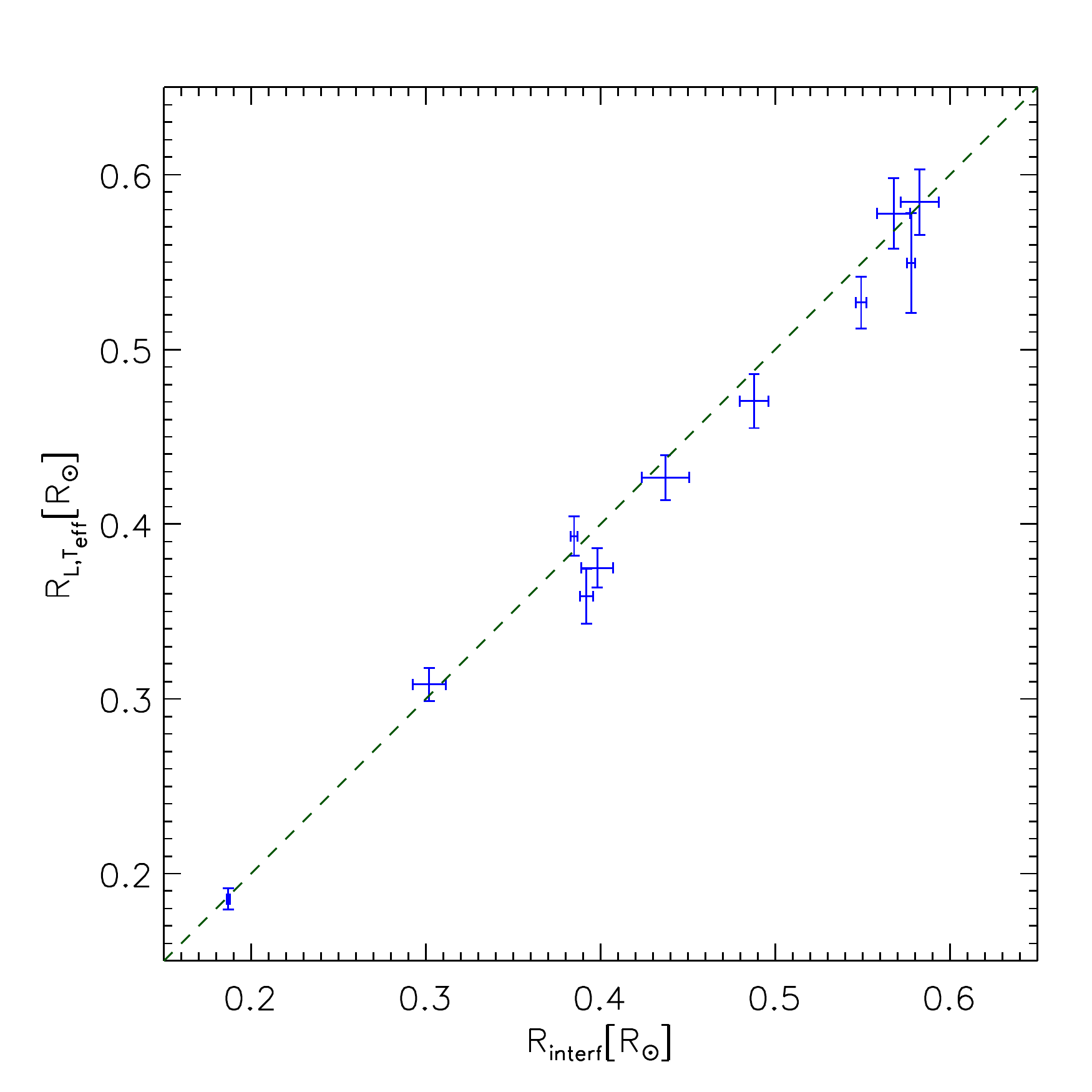}
   \caption{Radii determined using the spectroscopic \teff\ from Section~\ref{sec:temperatures},
            the bolometric luminosity from Section~\ref{sec:luminosities}, and
            Stefan-Boltzmann's law plotted against interferometric radii from \citet{2012ApJ...757..112B}.
            The two discussed outliers are Gl~411 and Gl~412A with $R_{\rm interf}$ of 0.392 and 0.398 $R\subsun$, respectively.
           }
      \label{fig:radii}
\end{figure}

Using the effective temperature \teff\ and luminosity $L$ from Section~\ref{sec:luminosities}
we calculated the radius $R$ using
Stefan-Boltzmann's law
\begin{equation}
\label{eq:rstefanb}
R=\left(\frac{L}{4\pi  \sigma \teff^4}\right)^{1/2}.
\end{equation}
The error for $R$ is then described by
\begin{equation}
\label{eq:drstefanb}
\left(\frac{\Delta R}{R}\right)^2  =  \left(2\,\frac{\Delta T_{\rm eff}}{T_{\rm eff}}\right)^2 + \left(\frac{1}{2}\frac{\Delta L}{L}\right)^2.
\end{equation}
When explicitly expressing $L$ by the distance $d$ and the
photometrically measured flux $S$ via \hbox{$L=4\pi d^2\cdot S$} the radius becomes
\begin{equation}
\label{eq:rstefanbdist}
R=\left(\frac{S}{\sigma \teff^4}\right)^{1/2}d
\end{equation}
and thus
\begin{equation}
\label{eq:drstefanbdist}
\left(\frac{\Delta R}{R}\right)^2  = \left(2\,\frac{\Delta T_{\rm eff}}{T_{\rm eff}}\right)^2 + \left(\frac{1}{2}\frac{\Delta S}{S}\right)^2 + \left(\frac{\Delta d}{d}\right)^2.
\end{equation}

In the era of satellite-based parallaxes, the relative error in distance $\Delta d/d$ is negligible
except for the very few stars without {\em Gaia} DR2 or {\em Hipparcos} parallaxes.
Therefore, the main contributions to the relative error arise from
the precision of the measured effective temperature,
$\Delta T_{\rm eff} / T_{\rm eff}$, and photometry $\Delta S / S$,
which are both typically between 1 and 2\,\%,
yielding typical $\Delta R/R$ between 2 and 3\,\%.

Our sample has 11 stars in common with the sample of stars with interferometric
radii presented by \citet{2012ApJ...757..112B}.
We used the interferometric angular diameter corrected for limb darkening $\Theta_{\rm LD}$
given in \citet{2012ApJ...757..112B}, but use {\em Gaia} DR2 distances $d$ when available
to calculate an interferometric radius $R_{\rm interf}$ via
\begin{equation}
\label{eq:rinterf}
R_{\rm interf} = \frac{1}{2} \Theta_{\rm LD} d.
\end{equation}
These  $R_{\rm interf}$ are listed in Table~\ref{tab:rinterf} along
with the angular diameters and distances that we used.
When we compared our spectroscopic radii with $R_{\rm interf}$ we find good agreement for 9 of the common stars, as can be seen in Fig.~\ref{fig:radii}.
Six of our radii have error bars comparable to the error bars
obtained from interferometry. For the remaining 5 stars, interferometry is
more precise, as expected.
Only the radii of the 2 targets Gl~411 (J11033+359) and Gl~412A (J11054+435), which are based on {\em Hipparcos} parallaxes,
differ significantly beyond their respective error bars (see Section~\ref{sec:outliers}).
The parallactic distances for the other 9 M~dwarfs are based on {\em Gaia} DR2 data.

\subsection{Masses (${\mathcal M}_{{\mathcal M}-R}$)}
\label{sec:mmr}

\begin{figure}
\centering
\includegraphics[width=1.0\hsize]{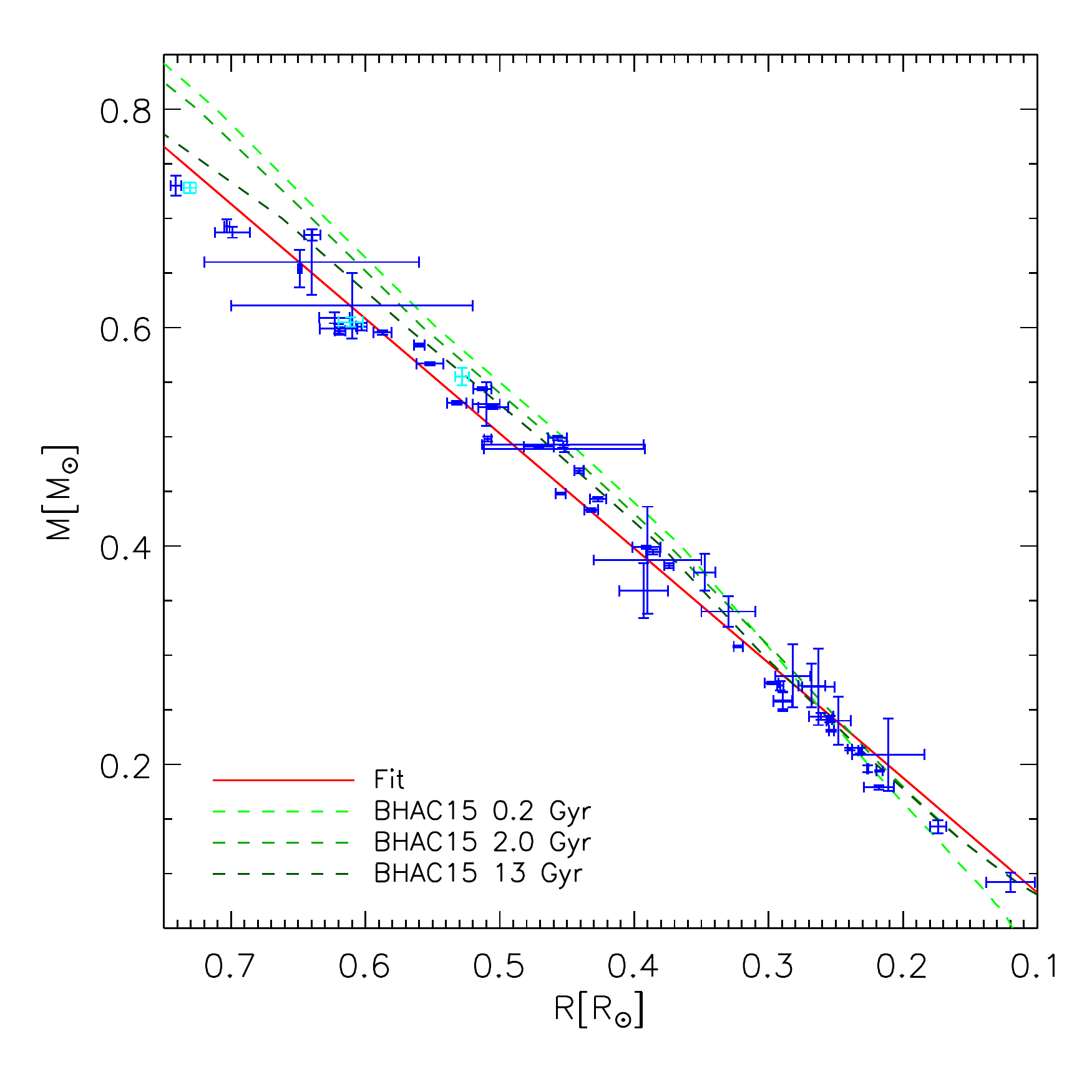}
   \caption{Masses and radii for the eclipsing binaries of Table~\ref{tab:mr}. Blue crosses indicate field stars,
            while turquoise crosses indicate,  from left to right, the low-metallicity globular cluster members
            NGC 6362 V41\,B \citep{2015AJ....150..155K}, M4 V65\,B \citep{2013AJ....145...43K},
            and   M55 V54\,B \citep{2014AcA....64...11K}.
            The red solid line shows our best linear fit.
            The increasingly darker green dashed lines represent isochrones for 0.2, 2, and 13\,Gyr
            from \citetalias{2015A&A...577A..42B}.
           }
      \label{fig:mr}
\end{figure}

We compiled a sample of 55 detached, double-lined, double-eclipsing, main-sequence M~dwarf
binaries from the literature (see Table~\ref{tab:mr} and Fig.~\ref{fig:mr}).
We excluded pre-main-sequence stars in very young open clusters and
very old stars in globular clusters.
From the published masses and radii of these eclipsing binaries, we
estimated the best linear fit \citep[following][]{Press:2007:NRE:1403886} of the mass-radius relation and found
for the radius and mass range given by Table~\ref{tab:mr},
\begin{equation}
\label{eq:mmr}
\frac{{\mathcal M}}{{\mathcal M}\subsun} =   (-0.0240\pm 0.0076) +      (1.055\pm 0.017) \cdot \frac{R}{R\subsun},
\end{equation}
or equivalently
\begin{equation}
\label{eq:mrm}
\frac{R}{R\subsun} =   (0.0282\pm0.0068) +      (0.935\pm0.015) \cdot \frac{{\mathcal M}}{{\mathcal M}\subsun}.
\end{equation}
Both fits have a non-reduced $\chi^2\approx 0.02$ translating into an rms~$\approx 0.02$ in solar
units.

By excluding members of young clusters,
the mass-radius relation
is well defined for M~dwarfs older than a few hundred million years.
This is underlined by the comparison with mass-radius relations for ages between 0.2 and 13\,Gyr from
evolutionary models for solar metallicity by \citet[][hereafter BHAC15]{2015A&A...577A..42B}, also shown in Fig.~\ref{fig:mr}.
            The explanation of the small difference between models and observations is beyond the scope of this
            paper. However, the small spread between the isochrones demonstrates that our
            linear mass-radius relation can be treated independently of age.
Furthermore, our sample in  Table~\ref{tab:mr} includes all known eclipsing binaries in the field,
independent of metallicity. Yet, our mass-radius relation is well defined, and, therefore,
we conclude that it does not significantly
depend on metallicity either. Such metallicity independence  of the mass-radius relation for
low-mass stars is also frequently found in evolutionary models \citep[see, e.g., Fig.~3 of][]{2018MNRAS.479.1953D}.
This is underlined by including three representative low-metallicity members of the open clusters
M\,4, M\,55, and NGC\,6362 in Fig.~\ref{fig:mr} (but not in the fits for Eqs.~\ref{eq:mmr} and \ref{eq:mrm}),
which match our mass-radius relation very well.

The reasons for the intrinsic spread in the observed mass-radius relation and the comparison
to theoretical models have been and are still being investigated elsewhere
\citep[e.g.,][]{2012ApJ...757...42F,2017AJ....154...30K,2018MNRAS.476...27T,2018MNRAS.481.1083P}.
As pointed out by \citet{2018MNRAS.481.1083P}, this spread sets a lower
limit on the error bars for the masses obtained with this method.
In any case, for this paper, we
assume that eclipsing binaries are representative of our single-star sample,
and we apply our mass-radius relation to derive masses.

In Table~\ref{tab:results} and for the remainder of this paper we label
masses determined with this method as ${\mathcal M}_{{\mathcal M}-R}$.
As already mentioned in the introduction, these are the masses that we have been using for the planet
discoveries of CARMENES. However, \citet{pre025} used the effective temperature from \citetalias{pre013}.
\citet{pre023} and \citet{pre022} additionally used the mass-radius relation
from \citet{msccasal} and the distances listed in {\em Gaia}~DR1 \citep{2016A&A...595A...2G}
or as measured by {\em Hipparcos} \citep{2007A&A...474..653V} for calculating the luminosity.
As a consequence, the masses we list for these three planet
hosts\footnote{J16167+672S (Gl~617A), J17578+046 (Gl~699), J1919+051N (Gl~752A)}
are about 5--8\% higher
than in the discovery papers. The largest contribution to this change is the
update of the parallax and, hence, luminosity.

\section{Comparison methods}
\label{sec:compare}

\subsection{Spectroscopic masses (${\mathcal M}_{\logg}$)}
\label{sec:mlogg}

If we use the surface gravity \logg\ from Section~\ref{sec:temperatures}, the mass is calculated via
\begin{equation}
\label{eq:mlogg}
{\mathcal M}=\frac{10^{\logg} R^2}{G},
\end{equation}
and its error is calculated via
\begin{equation}
\label{eq:dmlogg}
\left(\frac{ \Delta {\mathcal M}}{{\mathcal M}}\right)^2 = \left( \ln 10 \, \Delta \log g\right)^2  + \left(2\,\frac{\Delta R}{R}\right )^2.
\end{equation}
In practice,
the error in \logg\ dominates the error in ${\mathcal M}$.
The precision of our radii $\Delta R/R$ is about 2 to 3\,\% (cf. Section~\ref{sec:radii}).
It would require $\Delta \logg \lesssim 0.03$\,dex
for both contributions to Eq.~\ref{eq:dmlogg} to
become of similar size.
However, our current $\Delta \logg=0.07$\,dex is about
twice as large.
In Table~\ref{tab:results} and for the remainder of this paper we label
the mass determined with this method as ${\mathcal M}_{\logg}$.

\begin{figure}
\centering
\includegraphics[width=\hsize]{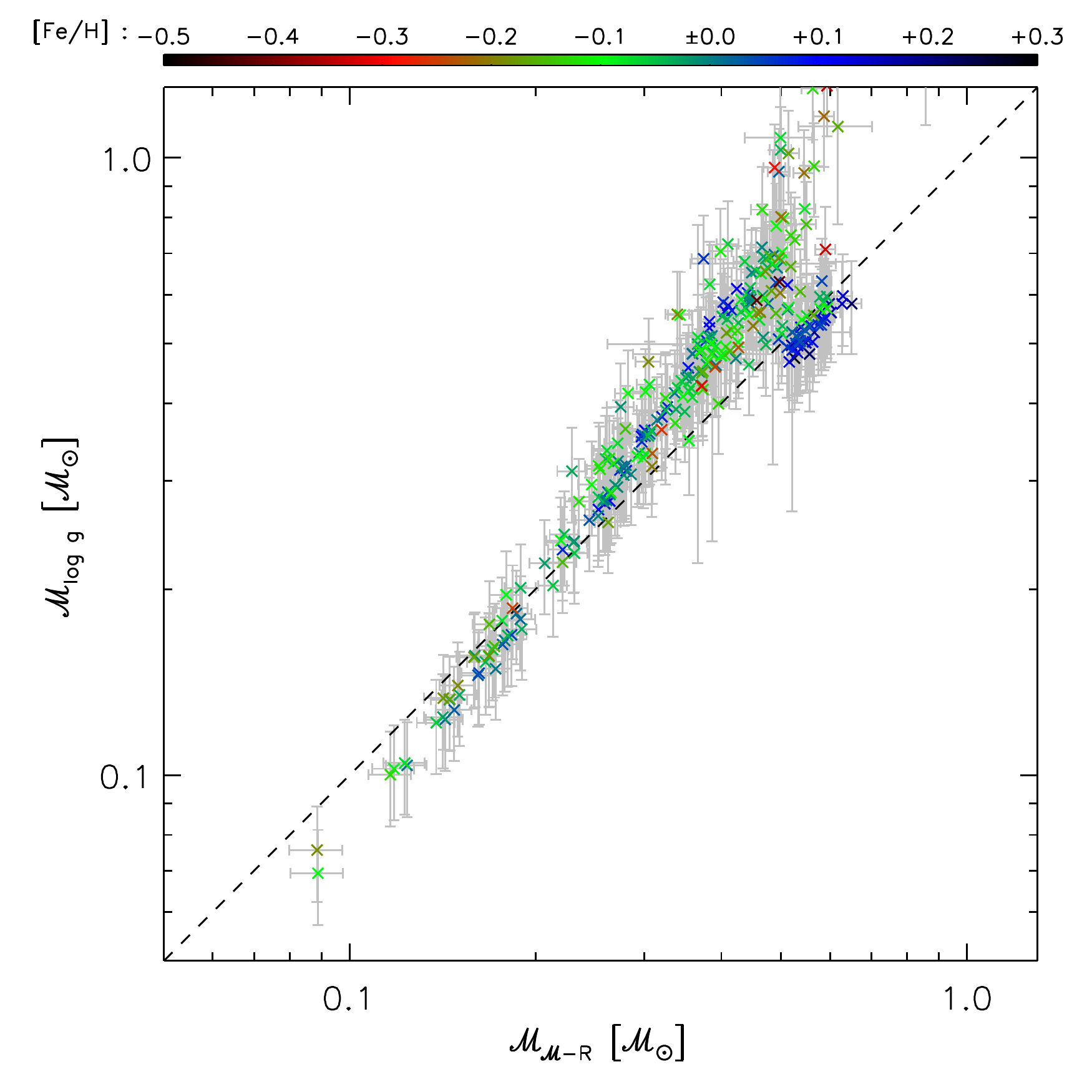}
   \caption{Comparison of masses for all stars in Table~\ref{tab:results} using
            our two methods described in Section~\ref{sec:mlogg} on the y-axis and Section~\ref{sec:mmr}
            on the x-axis.
            See Fig.~\ref{fig:mmr_mlogg_wo} for the same plot without error bars
            and linear axes.
           }
      \label{fig:mmr_mlogg}
\end{figure}

We compare ${\mathcal M}_{{\mathcal M}-R}$ from above with ${\mathcal M}_{\logg}$ 
in Fig.~\ref{fig:mmr_mlogg}. We find systematically lower ${\mathcal M}_{\logg}$~values for the
lowest (${\mathcal M}_{{\mathcal M}-R}\lesssim0.2{\mathcal M}\subsun$) and
highest (${\mathcal M}_{{\mathcal M}-R}\gtrsim 0.5{\mathcal M}\subsun$) ${\mathcal M}_{{\mathcal M}-R}$~values.
In between, the  ${\mathcal M}_{\logg}$~values are higher than the ${\mathcal M}_{{\mathcal M}-R}$~values,
and in particular for ${\mathcal M}_{{\mathcal M}-R}$ between 0.4 and 0.5 ${\mathcal M}\subsun$ there is a large
scatter without any correlation, except for the highest metallicities, which seem
to form a curve with a sharp turn around 0.5 ${\mathcal M}\subsun$.

\begin{figure}
\centering
\includegraphics[width=1.0\hsize]{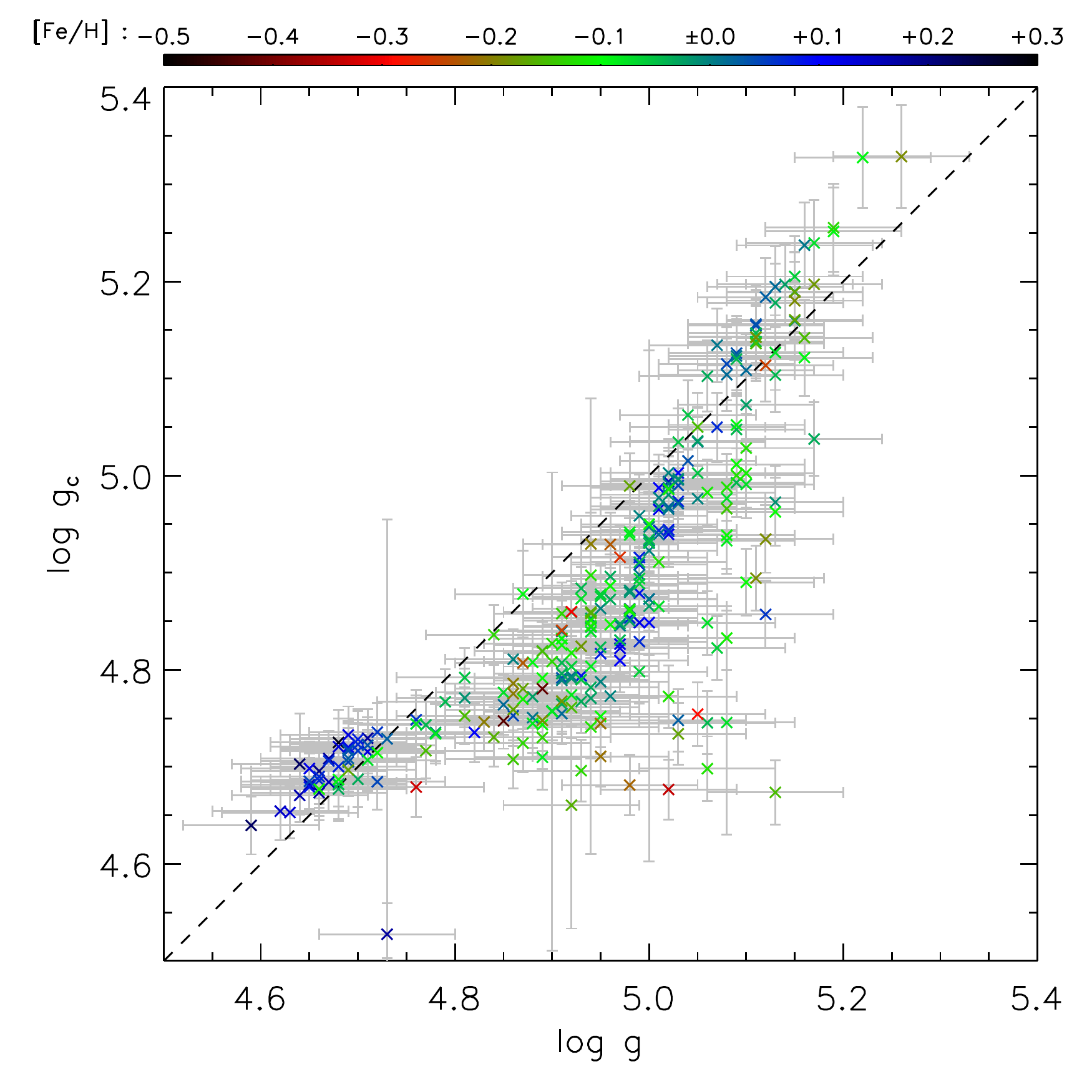}
   \caption{Value of $\logg_c$ calculated using ${\mathcal M}_{{\mathcal M}-R}$ from Section~\ref{sec:mmr} and using $R$ from Section~\ref{sec:radii}
            against the spectroscopic \logg\ from Section~\ref{sec:temperatures}.
           }
      \label{fig:logg}
\end{figure}

Furthermore, we reverse the calculation and obtain
the surface gravity  from ${\mathcal M}_{{\mathcal M}-R}$ and Stefan-Boltzmann's radius of Section~(\ref{sec:radii}).
We call this $\logg_c$ and list this value in Table~\ref{tab:results} as well.
When we compare the spectroscopic \logg\ from Section~\ref{sec:temperatures} to
this $\logg_c$ in Fig.~\ref{fig:logg} we find, as expected, a similar behavior
as above when comparing ${\mathcal M}_{{\mathcal M}-R}$ with ${\mathcal M}_{\logg}$:
there is a large scatter in the mid-value regime but there are consistent results 
for the highest and lowest surface gravities, i.e., the least and most massive stars.
And, again, the highest metallicity stars form a curve with less scatter than the other metallicities.
This is a similar result as found by \citetalias{pre013} in their Fig.~6, where some spectroscopic \logg\ values
deviate from literature values in a similar fashion.
Also similar to \citetalias{pre013} and as mentioned in Section~\ref{sec:temperatures},
our \logg\ is connected to \teff. This metallicity dependent connection
causes systematic effects due to the underlying evolutionary models.
In particular, the \logg\ values for high-metallicity results are obtained
by extrapolating from solar metallicity isochrones \citepalias{pre013}.

\begin{figure}
\centering
\includegraphics[width=1.0\hsize]{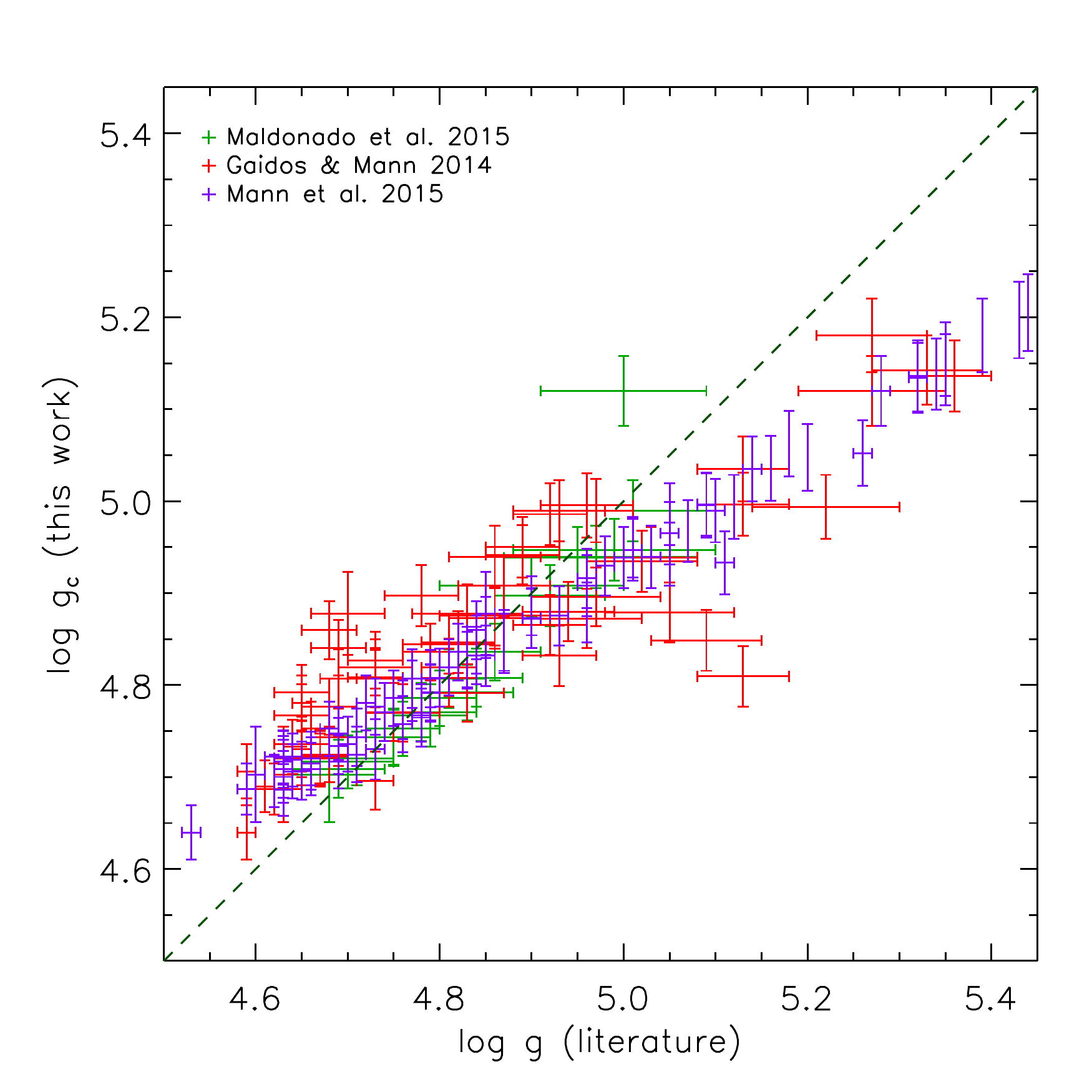}
   \caption{Value of $\logg_c$ calculated using ${\mathcal M}_{{\mathcal M}-R}$ from Section~\ref{sec:mmr} and using $R$ from Section~\ref{sec:radii}
against \logg\ from the literature.
           }
      \label{fig:logglit}
\end{figure}

To examine whether $\logg_c$ is more reliable than \logg, we compare our $\logg_c$
with literature values in Fig.~\ref{fig:logglit}
to produce a similar plot as Fig.~6 of \citetalias{pre013}.
When comparing Fig.~6 of \citetalias{pre013} to our Fig.~\ref{fig:logglit} we
find an improvement.
The values published by \citet{2015A&A...577A.132M} 
generally agree well with our $\logg_c$. The \logg\ values derived from radii and
masses published by \citet{2015ApJ...804...64M} show a linear correlation to ours,
although its slope is smaller than unity.
Only the  \logg\ values derived from radii and masses published by \citet{2014ApJ...791...54G} show
a very weak correlation to ours. Since the \logg\ values that have
recently been published by \citet{2018A&A...620A.180R} do not correlate with
\citetalias{pre013} or other literature values, we did not include these values in Fig.~\ref{fig:logglit}.
When we now use $\logg_c$ as an input for our spectral analysis described
in Section~\ref{sec:temperatures} and keep it fixed,
the resulting effective temperatures (and metallicities) do not change significantly and,
therefore, neither do the radii nor the masses determined in Section~\ref{sec:method}.

Despite the large error bars inherent in calculating ${\mathcal M}_{\logg}$,
and the problems described above, this method should not be generally dismissed.
With precise input data as we have here, the error bars of ${\mathcal M}_{\logg}$
become small enough so that the correlation in Fig.~\ref{fig:mmr_mlogg} becomes visible.
However, since the ${\mathcal M}_{{\mathcal M}-R}$ masses have a much smaller error bar
than the ${\mathcal M}_{\logg}$ masses we decided to keep using ${\mathcal M}_{{\mathcal M}-R}$.

\subsection{Photometric masses (${\mathcal M}_{{\mathcal M}-K_s}$)}
\label{sec:kband}

For M~dwarfs well established mass-magnitude relations exist for
infrared filters determined by \citet{2000A&A...364..217D}, \citet{2016AJ....152..141B},
and most recently by \citet{2018arXiv181106938M}.
The latter provided two variants: the first is independent of
metallicity and the second one accounts for metallicities $-0.4<{\rm [Fe/H]}<+0.3$
when using their conservative validity range for metallicity.
Both are valid for $4.5{\rm\,mag} < M_{Ks} < 10.5$\,mag, where $M_{Ks}$ is the absolute magnitude in the 2MASS $K_s$ band.
Since they also compared their results extensively to the well-established \citep{2015ApJ...804...64M}
${\mathcal M}-M_K$ relations of \citet{2000A&A...364..217D} and \citet{2016AJ....152..141B} and found
good (but not perfect, see below) agreement, and since we determined
metallicities (Section~\ref{sec:temperatures}) within the recommended range,
we employed the metallicity dependent variant of \citet{2018arXiv181106938M}
by using their recommended software\footnote{\tt \url{https://github.com/awmann/M\_-M\_K-}}.
However, the metallicity independent relation yields only marginally
different numbers.
We list the resulting masses, ${\mathcal M}_{{\mathcal M}-K_s}$, in Table~\ref{tab:results}.

\begin{figure}
\centering
\includegraphics[width=\hsize]{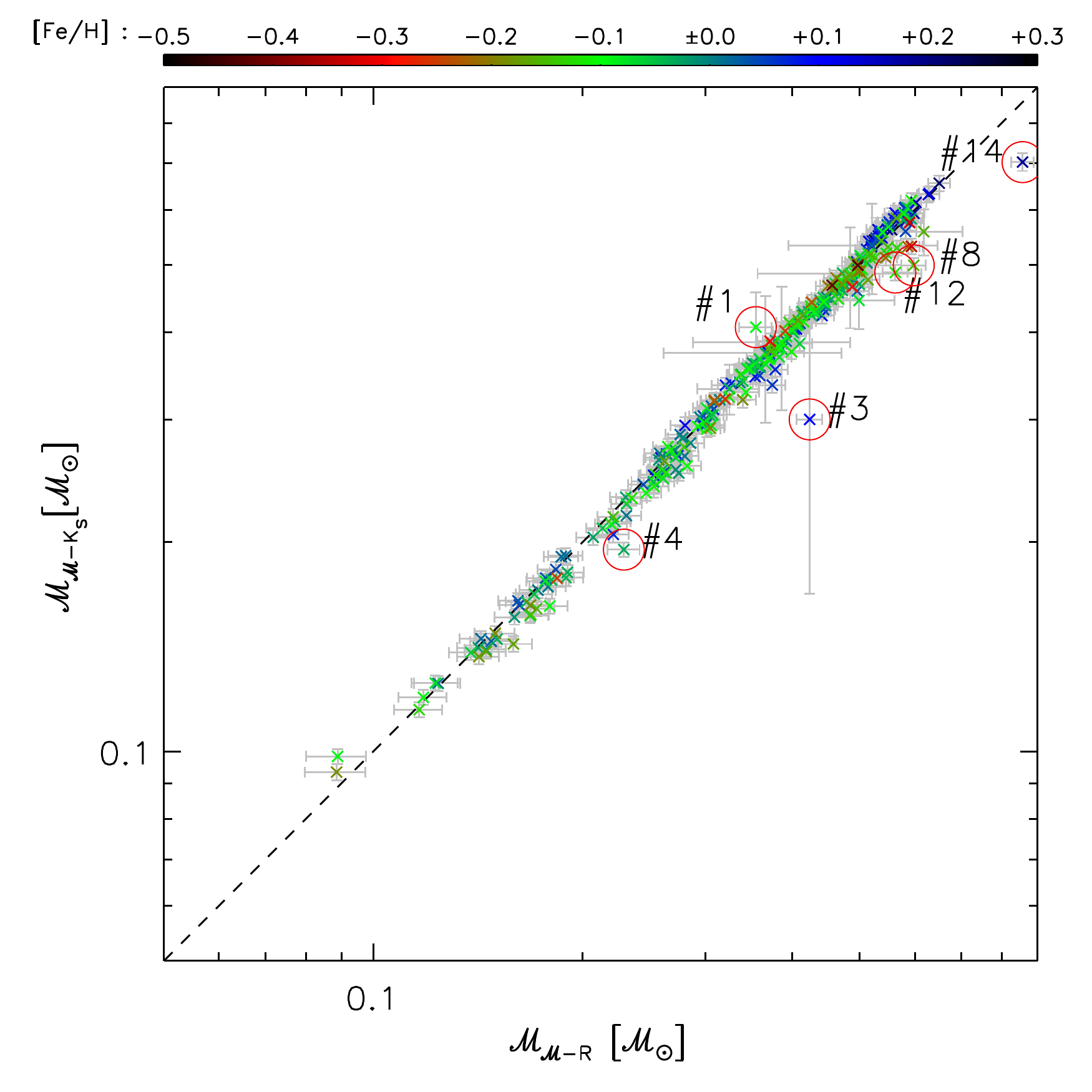}
  \caption{Comparison of masses for all stars in Table~\ref{tab:results} using the method from
            \citet{2018arXiv181106938M} as described in Section~\ref{sec:kband} on the y-axis
            and
            the method described in Section~\ref{sec:mmr} on the x-axis.
            The metallicities are color coded as indicated.
            \protect Obvious outliers, and stars discussed in reference to this figure in Section~\ref{sec:outliers}, are encircled:
\#1=Gl~411~(J11033+359), \#3=GJ~4063~(J18346+401), \#4=GJ~1235~(J19216+208), \#8=1RXS~J050156.7+010845~(J05019+011), \#12=StKM~2-809~(J12156+526), and \#14=K2-33~(J16102-193).
             See Fig.~\ref{fig:mmr_rel} for a different comparison between the plotted values.
           }
      \label{fig:mmr}
\end{figure}

A comparison between our ${\mathcal M}_{{\mathcal M}-R}$ masses and the photometrically determined masses ${\mathcal M}_{{\mathcal M}-K_s}$ is
given in Fig.~\ref{fig:mmr}.
Except for individual outliers (which is discussed in Section~\ref{sec:outliers}), the two
mass values agree very well.
When we, alternatively, compare our  ${\mathcal M}_{{\mathcal M}-R}$ values with the masses
obtained from the two relations of \citet{2000A&A...364..217D} and \citet{2016AJ....152..141B},
we find disagreement of just beyond the size of
our error bars for ${\mathcal M}_{{\mathcal M}-R}\gtrsim0.3{\mathcal M}\subsun$
but agreement otherwise.
This is the same result described by \citet{2018arXiv181106938M}, who found discrepancies
of about 10\% between their values and the values obtained from the two previous relations above $0.3{\mathcal M}\subsun$.
For the remainder of this paper we use ${\mathcal M}_{{\mathcal M}-R}$ since
differences between ${\mathcal M}_{{\mathcal M}-R}$  and ${\mathcal M}_{{\mathcal M}-K_s}$
are small.

As mentioned before, \citet{pre015} in their planet analysis and \citet{pre021} in their sample
overview based their stellar masses on a mass-magnitude relation, but
used a combination of the \citet{2000A&A...364..217D} and \citet{2016AJ....152..141B} relations
together with {\em Gaia} DR1 parallaxes.
Therefore, our ${\mathcal M}_{{\mathcal M}-K_s}$ masses
differ accordingly.

\subsection{PARSEC-based masses (${\mathcal M}_P$)}
\label{sec:padova}

Another method we used consists of a Bayesian approach applied to the PARSEC library of stellar
evolution models to infer stellar parameters, as performed by \citet{2018MNRAS.479.1953D}.
We applied this technique to 262 targets from Table~\ref{tab:results}, assuming
solar metallicity, and with data for the SDSS $r$ band, the 2MASS $J$ band,
and {\em Gaia} DR2 parallaxes to convert the apparent magnitudes $J$ into $M_J$.
The input parameters to feed the method were then $M_J$, $r-J$, and [Fe/H]~=~0.0$\pm$0.2\,dex.
The uncertainty in the iron-to-hydrogen ratio is reasonable since 
it reflects the range of [Fe/H] that we found in Sec.~\ref{sec:temperatures}
\citepalias[see also Fig.~5 of ][]{pre013}, and most of our targets belong to the
thin disc \citep{phdmiriam} for which this is a typical range.

A grid of PARSEC isochrones
\citep[version 1.2S;][]{2012MNRAS.427..127B,2014MNRAS.444.2525C,2015MNRAS.452.1068C,2014MNRAS.445.4287T}
were downloaded
and arranged, where [Fe/H] ranges from $-$2.18 to 0.47 in steps of 0.05 dex, age
spans from 2 Myr to 13.1\,Gyr with steps of 0.1\%, and the initial mass
runs from 0.09 ${\mathcal M}\subsun$ to the highest mass established by the corresponding
stellar lifetime.

This method yields mass, radius, luminosity, metallicity, age, and all derivable parameters.
For this work we only use mass and radius and list these in Table~\ref{tab:results} as $R_P$
and ${\mathcal M}_P$. The remaining results will be presented and discussed in a separate publication.

\begin{figure}
\centering
\includegraphics[width=\hsize]{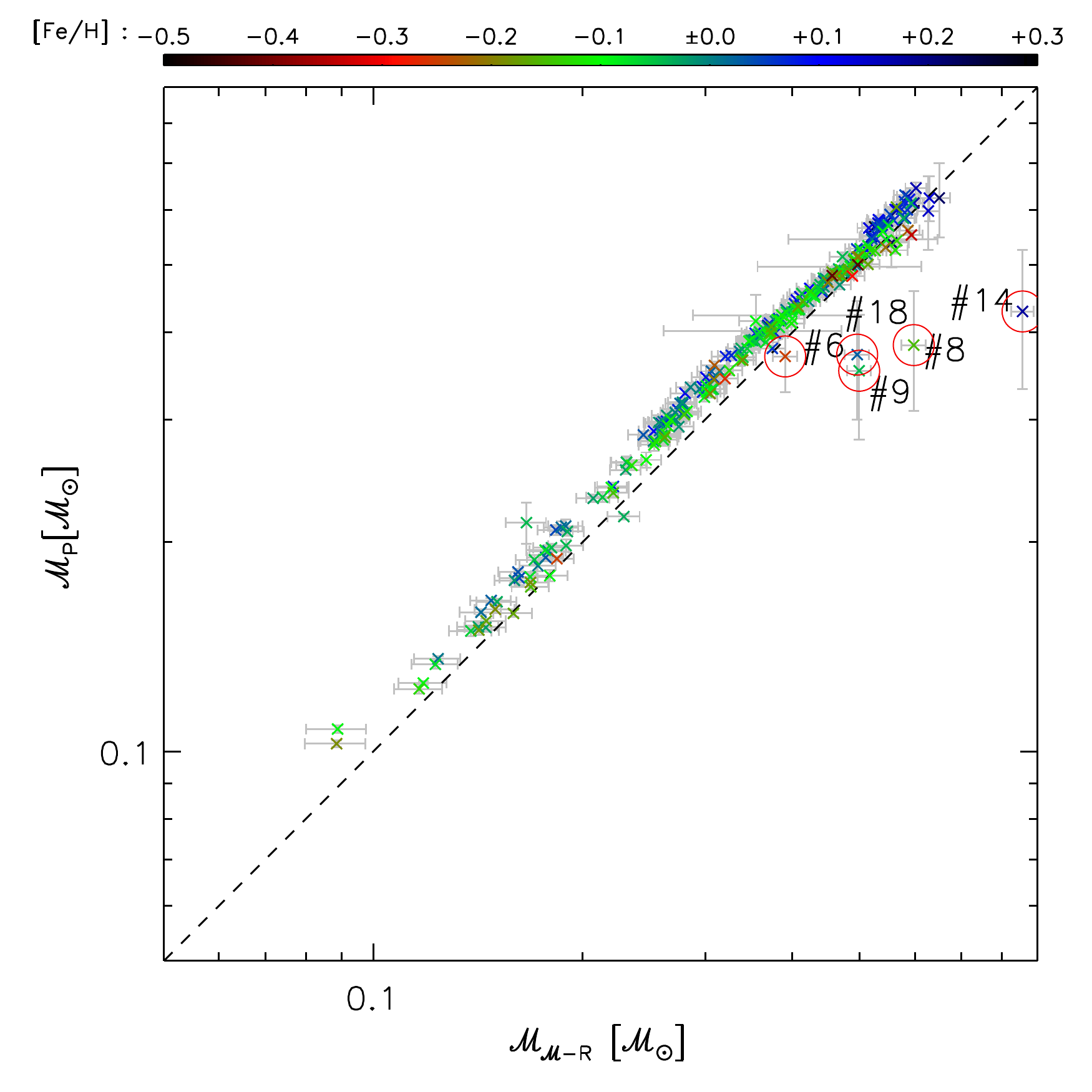}
  \caption{Comparison of masses for all stars in Table~\ref{tab:results} using
            the method described in Section~\ref{sec:padova} on the y-axis
            and
            the method described in Section~\ref{sec:mmr} on the x-axis.
            The metallicities are color coded as indicated.
            \protect Obvious outliers, and stars  discussed in reference to this figure in Section~\ref{sec:outliers}, are encircled:
\#6=Gl~15A~(J00183+440), \#8=1RXS~J050156.7+010845~(J05019+011), \#9=RX~J0506.2+0439~(J05062+046), \#14=K2-33~(J16102-193), and \#18=RBS~365~(J02519+224).
            }
      \label{fig:mpadova}
\end{figure}

When we compare ${\mathcal M}_{{\mathcal M}-R}$ with ${\mathcal M}_P$ in Fig.~\ref{fig:mpadova}
we find good correlation except for five outliers.
They are the very young objects K2-33 (J16102-193), 1RXS J050156.7+010845 (J05019+011,) and RX J0506.2+0439 (J05062+046),
and the two stars RBS~365 (J02519+224) and  Gl~15A (J00183+440).
All of these targets are discussed in Section~\ref{sec:outliers} below.
However, all ${\mathcal M}_P$ values are systematically slightly higher than ${\mathcal M}_{{\mathcal M}-R}$.
This means that the ${\mathcal M}_P$~values are also systematically offset to
the ${\mathcal M}_{{\mathcal M}-K_s}$~values in the same way.

Once again, for the remainder of this paper we use ${\mathcal M}_{{\mathcal M}-R}$ since systematic
differences between ${\mathcal M}_{{\mathcal M}-R}$  and ${\mathcal M}_P$
do not change the overall picture, as described in Section~\ref{sec:properties}.

\section{Discussion}
\label{sec:discussion}

\subsection{Sample properties}
\label{sec:properties}

\begin{figure}
\centering
\includegraphics[width=\hsize]{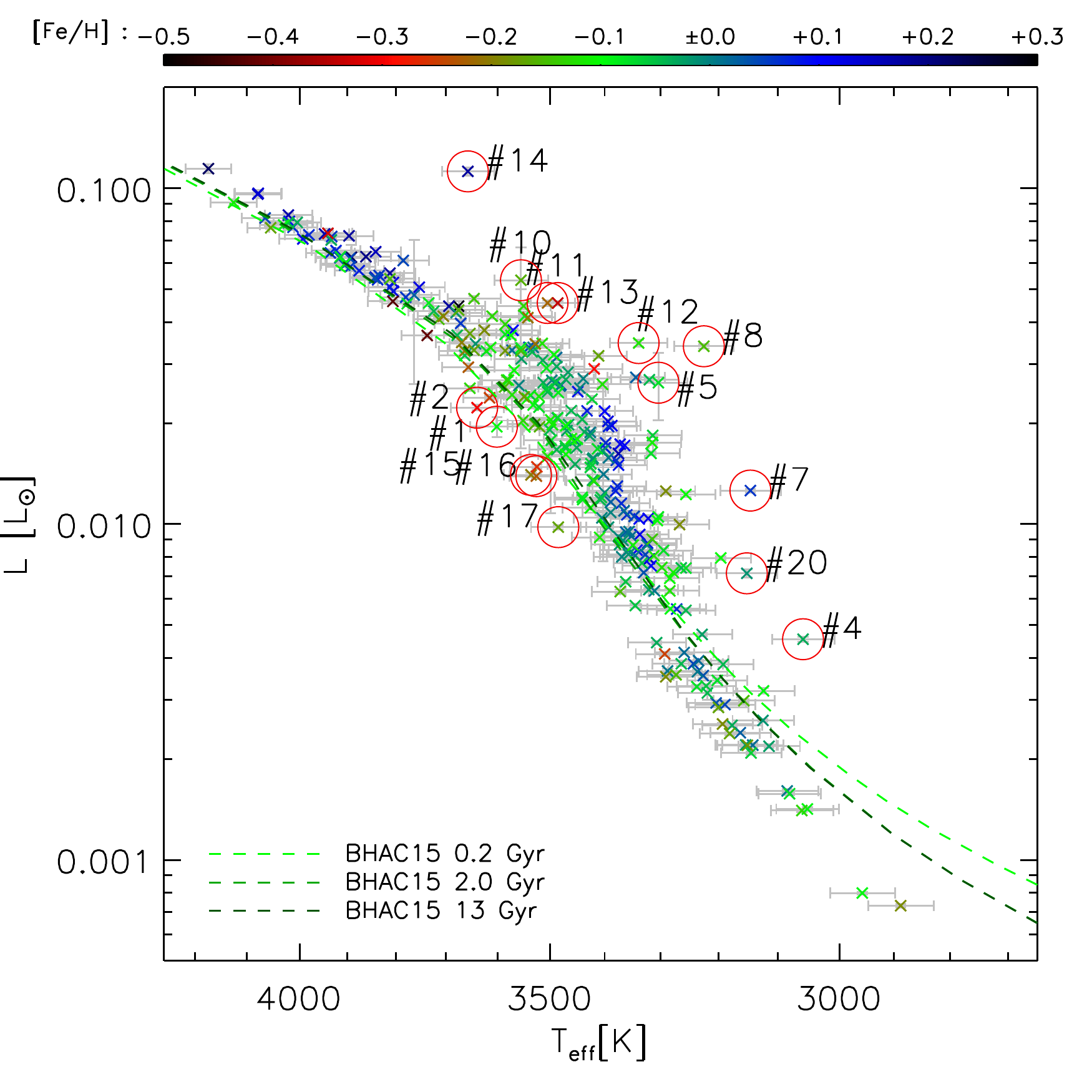}
   \caption{Hertzsprung-Russell diagram
             using $L$ from Section~\ref{sec:luminosities} and \teff\ from
             Section~\ref{sec:temperatures}.
             The metallicities are color coded as indicated.
            For comparison we plot isochrones for 0.2, 2, and 13\,Gyr (in increasingly darker green)
            from \citetalias{2015A&A...577A..42B}.
            \protect Obvious outliers, and stars discussed in reference to this figure in Section~\ref{sec:outliers}, are encircled:
\#1=Gl~411~(J11033+359), \#2=Gl~412A~(J11054+435), \#4=GJ~1235~(J19216+208), \#5=2M~J06572616+7405265~(J06574+740), \#7=RX~J0447.2+2038~(J04472+206), \#8=1RXS~J050156.7+010845~(J05019+011), \#10=G~234-057~(J09133+688), \#11=TYC~3529-1437-1~(J18174+483), \#12=StKM~2-809~(J12156+526), \#13=GJ~9520~(J15218+209), \#14=K2-33~(J16102-193), \#15=Gl~745A~(J19070+208), \#16=Gl~745B~(J19072+208), \#17=Gl~465~(J12248-182), and \#20=LP~022-420~(J15499+796).
             Outliers \#12 and \#13 lie almost on top of each other.
            See Fig.~\ref{fig:hrdhalpha} for the same plot but using the pseudo equivalent width pEW$^\prime_{{\rm H}\alpha}$
            of \citet{pre034} for the color coding.
            See also Fig.~\ref{fig:hrd_rel} for a different comparison between the plotted values.
            }
      \label{fig:hrd}
\end{figure}

Having established a way to measure the masses of our targets, we investigate further statistical properties of our sample.
When we use our results and produce a Hertzsprung-Russell diagram (HRD)
in Fig.~\ref{fig:hrd},
we find all targets on the main sequence except for a few outliers. 
Most of these outliers are young. This is discussed on a case-by-case
basis below.

\begin{figure}
\centering
\includegraphics[width=\hsize]{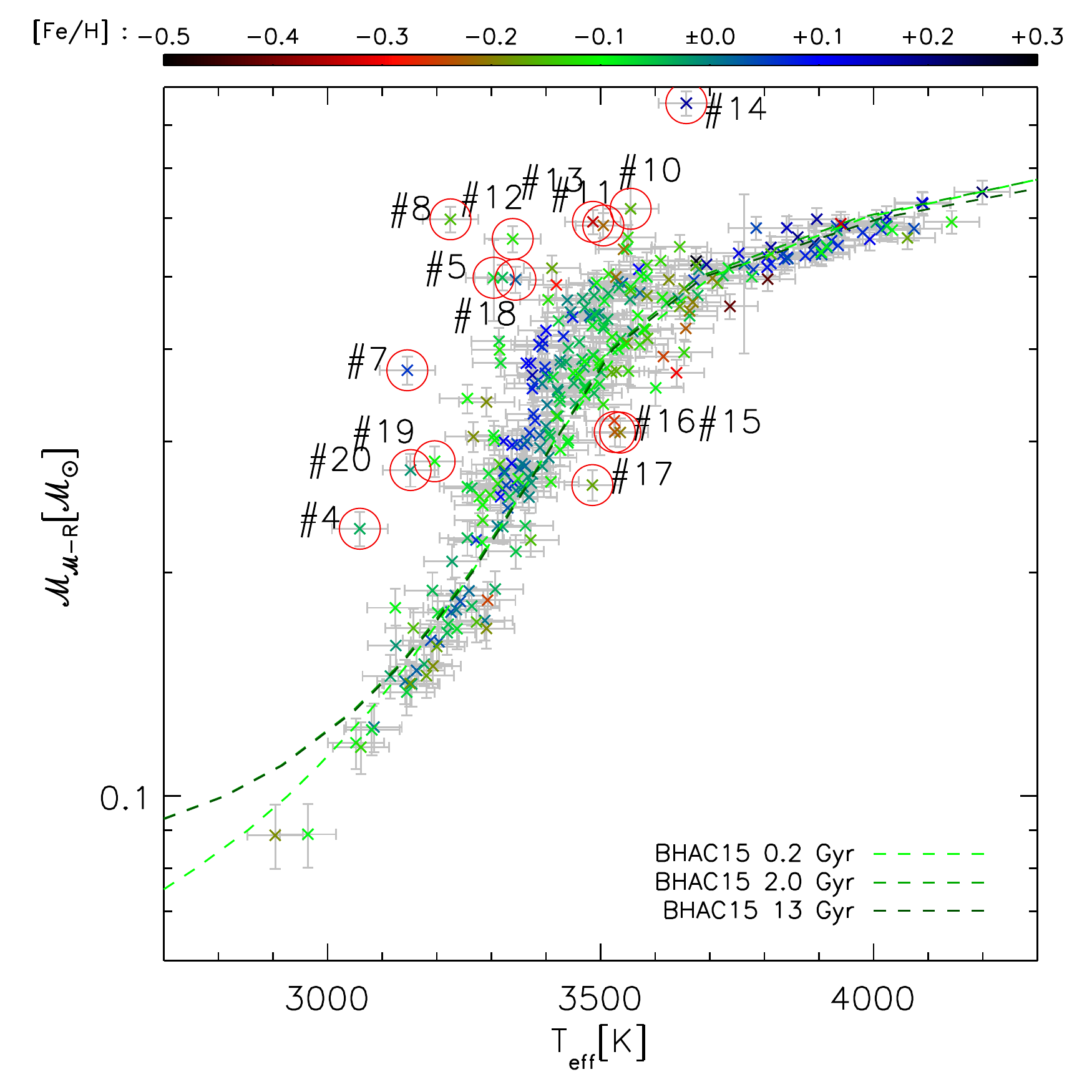}
   \caption{ \ Values of ${\mathcal M}_{{\mathcal M}-R}$ from Section~\ref{sec:mmr} plotted against \teff\ from Section~\ref{sec:temperatures}.
            The metallicities are color coded as indicated.
            For comparison we plot isochrones for 0.2, 2, and 13\,Gyr (in increasingly darker green)
            from \citetalias{2015A&A...577A..42B}.
            \protect Obvious outliers, and stars discussed in reference to
this figure in Section~\ref{sec:outliers}, are encircled:
\#4=GJ~1235~(J19216+208), \#5=2M~J06572616+7405265~(J06574+740), \#7=RX~J0447.2+2038~(J04472+206), \#8=1RXS~J050156.7+010845~(J05019+011), \#10=G~234-057~(J09133+688), \#11=TYC~3529-1437-1~(J18174+483), \#12=StKM~2-809~(J12156+526), \#13=GJ~9520~(J15218+209), \#14=K2-33~(J16102-193), \#15=Gl~745A~(J19070+208), \#16=Gl~745B~(J19072+208), \#17=Gl~465~(J12248-182), \#18=RBS~365~(J02519+224), \#19=RX~J1417.3+4525~(J14173+454), and \#20=LP~022-420~(J15499+796).
             Outliers \#12 and \#13 lie almost on top of each other.
           }
      \label{fig:mteff}
\end{figure}

\begin{figure}
\centering
\includegraphics[width=\hsize]{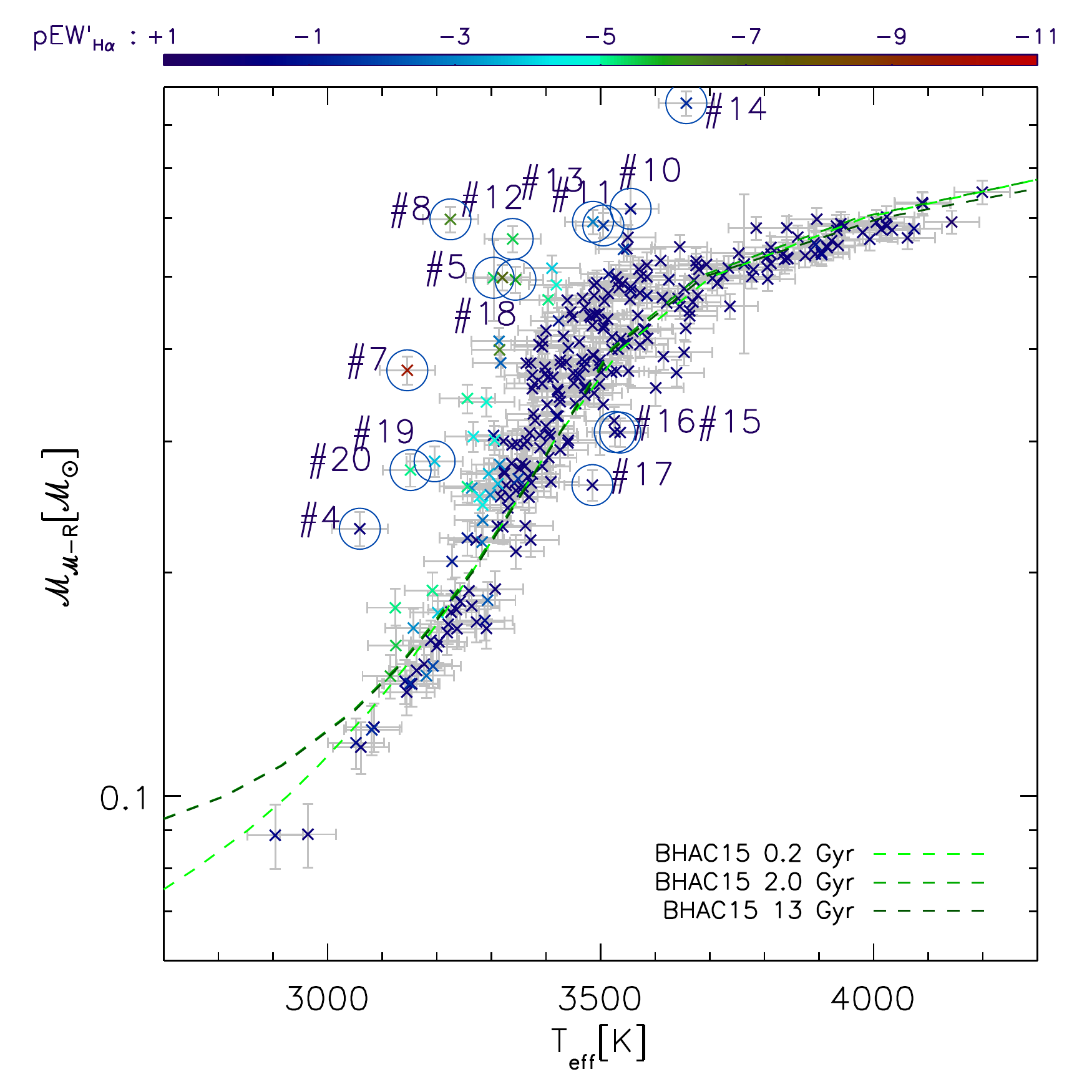}
   \caption{Same as Fig.~\ref{fig:mteff}, but using the pseudo equivalent width pEW$^\prime_{{\rm H}\alpha}$
            of \citet{pre034} for the color coding. Negative values of pEW$^\prime_{{\rm H}\alpha}$ denote H$\alpha$ emission.
            }
      \label{fig:mteffhalpha}
\end{figure}

Furthermore, when we plot \teff\ against the mass of the star we find the relation shown in Figs.~\ref{fig:mteff}
and Fig.~\ref{fig:mteffhalpha}.
While there is a large spread, the distribution is well consistent with the isochrones by \citetalias{2015A&A...577A..42B}.
Again, we can identify outliers, most of which are young or active, and all of these cases are discussed below.
While the HRD only compares our independently determined luminosities $L$ and effective temperatures \teff,
Fig.~\ref{fig:mteff} directly shows the results of our method described in Section~\ref{sec:mmr}.

\begin{figure}
\centering
\includegraphics[width=\hsize]{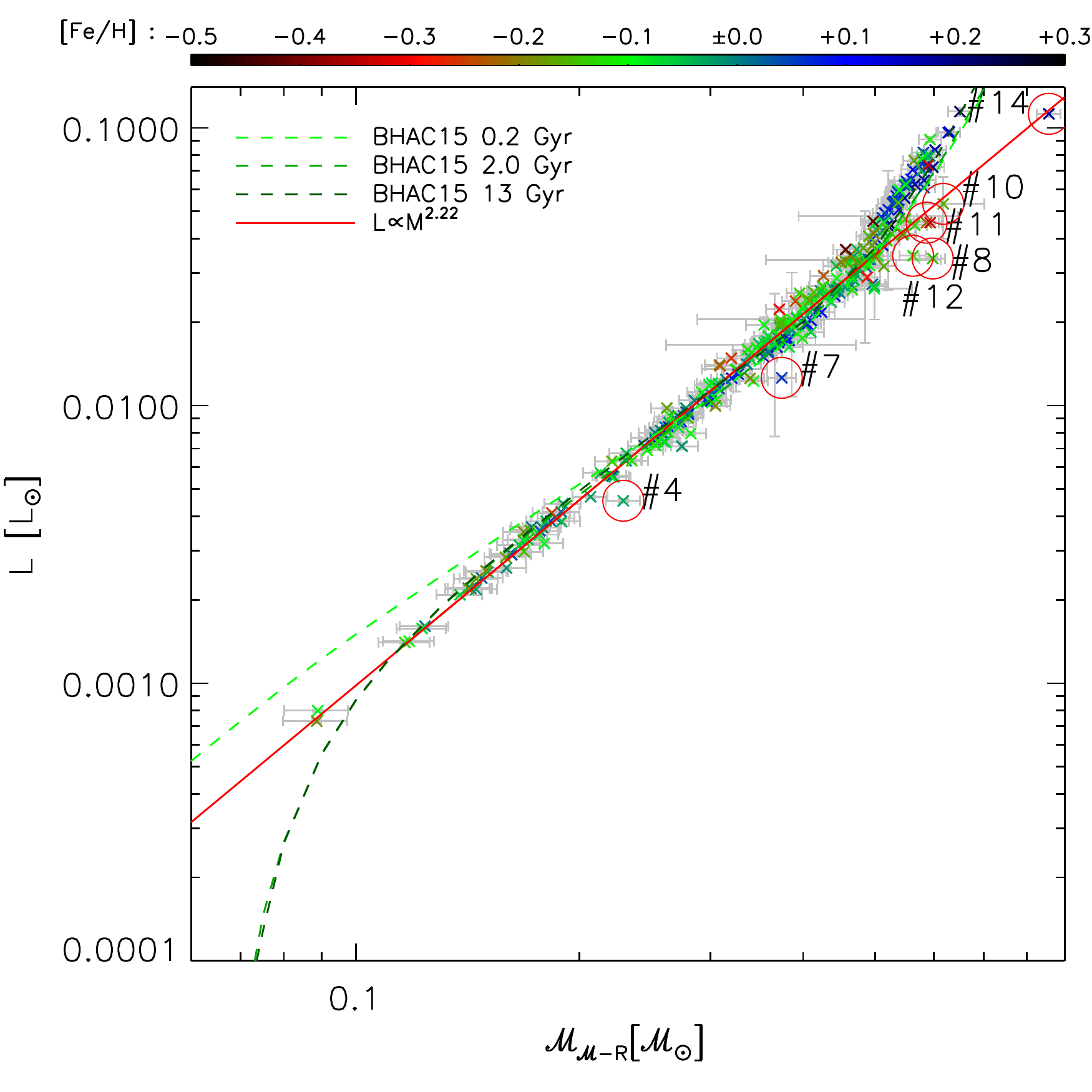}
   \caption{
             Mass-luminosity relation using $L$ from Section~\ref{sec:luminosities} and
             ${\mathcal M}_{{\mathcal M}-R}$ from Section~\ref{sec:mmr}.
            The metallicities are color coded as indicated.
            For comparison we plot isochrones for 0.2, 2, and 13\,Gyr (dashed lines in increasingly darker green)
            from \citetalias{2015A&A...577A..42B}.
            The red solid line indicates the fit we obtained in Eq.~\ref{eq:ml}
            \protect Obvious outliers, and stars discussed in reference to
this figure in Section~\ref{sec:outliers}, are encircled:
\#4=GJ~1235~(J19216+208), \#7=RX~J0447.2+2038~(J04472+206), \#8=1RXS~J050156.7+010845~(J05019+011), \#10=G~234-057~(J09133+688), \#11=TYC~3529-1437-1~(J18174+483), \#12=StKM~2-809~(J12156+526), and \#14=K2-33~(J16102-193).
            }
      \label{fig:lm}
\end{figure}

Finally, when we plot the bolometric luminosity from Section~\ref{sec:luminosities} against the mass of the star
to create a (bolometric) mass-luminosity relation we obtain Fig.~\ref{fig:lm}.
Again it is well consistent with the isochrones by \citetalias{2015A&A...577A..42B}.
When we fit a power law to these data for $0.1{\mathcal M}\,\subsun < {\mathcal M}_{{\mathcal M}-R} < 0.5{\mathcal M}\,\subsun$ we find a flat power law of
\begin{equation}
\label{eq:ml}
L\propto {\mathcal M}^{2.22\pm 0.02}
 ,
\end{equation}
which deviates from power laws for $L-{\mathcal M}$ relations that
average the whole main sequence yielding exponents of at least three.
However, this agrees with results obtained for very low-mass stars that
show a flattening of this power law at the bottom of the main sequence.
This flattening can be seen already in, for example,  \citet{2010A&ARv..18...67T} or \citet{2018MNRAS.479.1953D}.
As a recent example,
\citet[][and similarly the references therein]{2018MNRAS.479.5491E} have derived a
comparable exponent of 2.028$\pm$0.135 for 
$0.179{\mathcal M}\,\subsun < {\mathcal M} < 0.45{\mathcal M}\,\subsun$.
For ${\mathcal M}_{{\mathcal M}-R} \gtrsim 0.5{\mathcal M}\,\subsun$ the relation
in Fig.~\ref{fig:lm} clearly required a different exponent, but our high-mass range is too
small to fit a reliable second power law.

\subsection{Outliers}
\label{sec:outliers}

Most of the outliers discussed in this section are young objects. They either belong
to young associations or to young moving groups. Our ${\mathcal M}_{{\mathcal M}-R}$
values, however, are based on the assumption that the stars are older than a few
hundred million years (cf. Section~\ref{sec:mmr}). The ${\mathcal M}_{{\mathcal M}-R}$ values
that we list for these young objects should be treated accordingly. As \citetalias{pre013} have already concluded, their effective temperatures are generally
not affected by this assumption. Under the further assumption that
the luminosity is not affected much either, the radii we list
are not affected much by any assumptions regarding ages either. In other words,
in cases in which we list or show a mass that is too high,
we also derive a similarly large radius (cf. Eqs.~\ref{eq:mmr} and \ref{eq:mrm}),
which may be expected for young objects.

Another set of the targets listed below are outliers and at the
same time, some of their entries in at least one of the photometric catalogs are marked
as being of bad quality.
If such systematic errors were accounted for, our error bars
would be increased significantly.

\subsubsection{Data-based reasons}

We first eliminate all outliers from our discussion that have
data-based reasons for their designation as outliers because we do not wish to misidentify physical reasons.

\paragraph{\#1=Gl 411 (J11033+359) and \#2=Gl 412A (J11054+435)}

For these two stars we obtain different radii than measured interferometrically (cf. Section~\ref{sec:radii}).
Both are missing in the latest {\em Gaia} catalog, therefore we use parallaxes from
the {\em Hipparcos} measurements. However, when comparing Eqs.~\ref{eq:rstefanbdist} and \ref{eq:rinterf}
it is obvious that the distance does not cause a mismatch between the two different
methods of computing radii. We confirm that by using the {\em Gaia} distance of Gl~412B
for Gl~412A since both form a common proper motion pair. The values $R$ and $R_{\rm interf}$ both changed
slightly, but the data point in Fig.~\ref{fig:radii} only moved parallel to the 1-1-line,
as expected.

Both stars have very reliable measurements of their effective temperatures since their fits
described in Section~\ref{sec:temperatures} are of high quality. Furthermore, both \teff\
agree well with the recent measurements of \citet{2015ApJ...804...64M} and \citet{2018A&A...620A.180R}. Therefore, the reason
for the mismatch to the very reliable interferometric radii
must lie in the photometry. As a test, we took three stars from Table~\ref{tab:rinterf}
that are in {\em Gaia} DR2 (Gl~15A, Gl~880 and Gl~699) and for which we obtain
good results using  {\em Gaia} parallaxes and photometry (cf. Section~\ref{sec:radii}).
But for this test we ignored all the {\em Gaia} bands $B_P$, $G$, and $R_P$, and calculated the luminosities with
the remaining photometry. Using VOSA, we obtained different luminosities and, hence, radii
that no longer agreed with their interferometric counterparts. This is because the
{\em Gaia} filters anchor the SED very reliably in the optical while 2MASS and allWISE anchor the
SED in the infrared.
In particular Gl~15A, which has unreliable 2MASS photometry (see below), showed the largest change
in the test.

As a consequence of its uncertain luminosity, radius, and mass, Gl~411 also appears below
the main sequence in the HRD in Fig.~\ref{fig:hrd} and is a mild outlier in Fig.~\ref{fig:mmr}.
In addition to not having {\em Gaia} magnitudes, Gl~411 has also 2MASS and allWISE magnitudes of
poor quality because of its brightness. In both catalogs Gl~411 is listed to have magnitudes with low quality flags (DCD in 2MASS and UUAA in allWISE).
As described above, our error bars would have to be larger.

\paragraph{\#3=GJ 4063 (J18346+401)}
This is an outlier when comparing ${\mathcal M}_{{\mathcal M}-R}$ to
the absolute $K_s$ band magnitude (cf. Fig.~\ref{fig:mks}).
It is also one of the outliers when comparing ${\mathcal M}_{{\mathcal M}-R}$
with ${\mathcal M}_{{\mathcal M}-K_s}$ in Fig.~\ref{fig:mmr}.
The reason for the deviations is the poor $K_s$-band photometry as indicated
by the 2MASS quality flags (AAU).

\paragraph{\#4=GJ 1235 (J19216+208)}
This object is an outlier in the HRD (Fig.~\ref{fig:hrd}),
when plotting mass against effective temperature in Fig.~\ref{fig:mteff}
and
in the $L-{\mathcal M}_{{\mathcal M}-R}$ plot in Fig.~\ref{fig:lm}.
However, we cannot explain this except by pointing out that
this is a relatively faint, high proper motion star located in a crowded field where
photometric contamination by background sources cannot be excluded.

\paragraph{\#5=2MASS~J06572616+7405265 (J06574+740)}
This object does not have a space-based parallax. We use the value obtained
with the United States Naval Observatory Robotic Astrometric Telescope \citep{2016AJ....151..160F},
which is the only published trigonometric parallax and has a large relative error of 11\%. Despite its large
error bar, it is an outlier in the HRD (Fig.~\ref{fig:hrd}),
and when plotting mass against effective temperature in Fig.~\ref{fig:mteff}.
However, as an X-ray source \citep{2009ApJS..184..138H} this is an active star \citep{2015ApJ...798...41A, pre034},
but we attribute its deviation in the
diagrams instead to a wrong parallactic distance ($d$ = 26$\pm$3\,pc).
In the literature, there have been at least four additional distance
determinations from photometry and spectroscopy \citep{2011AJ....142..138L,2013AJ....145..102L,phdmiriam},
and they all point toward a closer
distance in the range $d$ = 14.0--19.6\,pc, which would lead to a
luminosity and mass consistent with its measured effective temperature.
We expect the next {\em Gaia} release to improve the precision of its luminosity
and to help clarify its nature.

\paragraph{\#6=Gl 15A (J00183+440)}
When plotting mass versus the $J$ magnitude (Fig.~\ref{fig:mj}) this object is a mild outlier. However,
it is also too bright in the infrared to have reliable 2MASS magnitudes as
indicated by their quality flags (DCE).

\subsubsection{Young age-based reasons}

\paragraph{\#7=RX J0447.2+2038 (J04472+206)}
This is one of the outliers in the HRD (Fig.~\ref{fig:hrd}) and when comparing
mass against \teff\ in Figs.~\ref{fig:mteff} and \ref{fig:mteffhalpha}.
It is very active \citep{2014AJ....147...70I,2018A&A...614A..76J,2018A&A...614A.122T},
and \citet{phdmiriam} identified this object
as a candidate member of the young disc population.

\paragraph{\#8=1RXS J050156.7+010845 (J05019+011) and \#9=RX J0506.2+0439 (J05062+046)}
The first is an outlier in all plots.
It was identified by \citet{2012AJ....144..109S} to be a member of the $\beta$~Pictoris
moving group and, hence, very young.
The second is a mild outlier and was identified by \citet{2012AJ....143...80S} as
a candidate member of $\beta$~Pictoris as well.
These are two of the outliers in Fig.~\ref{fig:mpadova} comparing ${\mathcal M}_P$
to ${\mathcal M}_{{\mathcal M}-R}$. Their corresponding masses differ by about 30\%.
For the former the PARSEC-based method (cf. Section~\ref{sec:padova})
yields for ${\mathcal M}_P=0.38\pm0.08{\mathcal M}\,\subsun$ an age of $\tau=36\pm16$~Myr.
For the latter the method yields for ${\mathcal M}_P=0.35\pm0.07{\mathcal M}\,\subsun$ an
age of $\tau=38\pm17$~Myr.
Both ages are similar to the currently assumed age of
$\beta$~Pictoris of $\tau=25\pm3$\,Myr \citep[e.g.,][]{2016A&A...596A..29M,2017AJ....154...69S}.

\paragraph{\#10=G~234-057 (J09133+688) and \#11=TYC~3529-1437-1 (J18174+483)}
Both stars are noticeable outliers in almost all plots.
However, both are young and candidate members of the AB Doradus association. The first was listed
as a candidate by \citet{2012AJ....143...80S}, and the second by \citet{2017ApJ...841...73K}.
The second is also an active star discussed in \citet{2018A&A...615A..14F}.

\paragraph{\#12=StKM 2-809 (J12156+526)}
This is an outlier in  the $L-{\mathcal M}_{{\mathcal M}-R}$ plot in Fig.~\ref{fig:lm} and 
the HRD (Fig.~\ref{fig:hrd}). But again, this is an active \citep{1986AJ.....92..139S,2018A&A...614A.122T},
young star, which was identified as a candidate member of the Ursa Major moving group \citep{phdmiriam}.

\paragraph{\#13=GJ 9520 (J15218+209)}
This is an outlier in almost every plot (e.g., in the HRD in Fig.~\ref{fig:hrd}), although not always an obvious one.
It is an active \citep{2018A&A...615A..14F} and young star belonging to the Local Association
\citep{2011MNRAS.410..190T}.

\paragraph{\#14=K2-33 (J16102-193)}
This is a pre-main-sequence star in the Upper Scorpius OB association. Its
age is estimated to be between 5 and 20~Myr \citep{2016Natur.534..658D,2016AJ....152...61M}.
It is a very obvious outlier in almost all plots. We measure unrealistic masses
above $0.8\,{\mathcal M}\subsun$ with all methods except that for obtaining  ${\mathcal M}_P$
described in Section~\ref{sec:padova}.
Therefore, it appears as one of the outliers in Fig.~\ref{fig:mpadova}.
The PARSEC-based method yields ${\mathcal M}_P=0.43\pm0.10\,{\mathcal M}\subsun$ and at the same time an age of
$\tau=8.0\pm3.5$~Myr.
Furthermore, it is by far the most distant target, and at
a distance of $d\approx140$\,pc located in Upper Scorpius, extinction
can no longer be neglected.

\subsubsection{Examples of possible surprising results}

While the stars discussed in this section are not
obvious outliers, they demonstrate certain
physical effects that can result in unexpected
masses or radii. Furthermore, these stars are only
examples and not a comprehensive list.
In particular the plots involving \teff\ (Figs.~\ref{fig:hrd}, \ref{fig:mteff} and \ref{fig:mteffhalpha})
show a natural spread \citepalias[as already described by][]{pre013},
and the decision on which star is an outlier in these plots would require
a non-trivial definition of a criterion.

\paragraph{\#15=Gl 745A (J19070+208) and \#16=Gl 745B (J19072+208)}
Both components of the wide binary system LDS~1017 are slightly below the main sequence
in the HRD (Fig.~\ref{fig:hrd}).
We also measure high effective temperatures for their derived masses
when comparing these two parameters in Fig.~\ref{fig:mteff}. Together with
very low activity they show a behavior akin to subdwarfs. Our derived [Fe/H]~$= -0.20\pm0.16$\,dex and [Fe/H]~$= -0.23\pm0.16$\,dex, respectively,
also point toward low-metallicity objects. Furthermore, \citet{2015ApJ...804...64M}
derived for Gl~745A an even lower metallicity of [Fe/H]~=~$-0.33\pm0.08$\,dex,
and \citet{2008MNRAS.390.1081H} already classified both components as subdwarfs.

\paragraph{\#17=Gl 465 (J12248-182)}
As the previous two stars, this isolated star is below the main sequence in
the HRD and has a very high effective temperature for its derived mass (see Fig.~\ref{fig:mteff}).
We measure [Fe/H]=$-$0.17$\pm$0.16\,dex for this object.
This metallicity is still too high to call it a subdwarf
as it is still consistent with the solar.
However, \citet{2015A&A...577A.132M}  measured 
an even lower metallicity of [Fe/H]~=~$-0.33\pm0.09$\,dex for this star.

\paragraph{\#18=RBS 365 (J02519+224), \#19=RX J1417.3+4525 (J14173+454) and \#20=LP 022-420 (J15499+796)}
These three stars are mild outliers in the HRD (Fig.~\ref{fig:hrd}) or when plotting mass
against \teff\ (Figs.~\ref{fig:mteff} and \ref{fig:mteffhalpha}). These, however,
are active stars that have, for instance, $\log L_{\rm H\alpha}/L_{\rm bol}>-4$ \citep{pre034}.
The latter two are also in the CARMENES radial-velocity-loud sample discussed by \citet{2018A&A...614A.122T}.

\subsection{How do the masses differ?}
\label{sec:masses}

Overall, all methods in determining masses yield consistent results, even
when using \logg\ in order to obtain ${\mathcal M}_{\logg}$.

The largest amount of information is used in the construction of ${\mathcal M}_{{\mathcal M}-R}$
(Section~\ref{sec:mmr})
and in the construction of ${\mathcal M}_{\logg}$ (Section~\ref{sec:mlogg}). The latter does not even use any empirical
relation, however, it yields the largest error bars.
Both ${\mathcal M}_{{\mathcal M}-R}$ and ${\mathcal M}_{{\mathcal M}-K_s}$ (Section~\ref{sec:kband}) use 
an empirical relation. However, the ${\mathcal M}-R$ relation is more fundamental
than the mass-magnitude relation. Mass and radius are fundamentally connected
via the hydrostatics of a star, whereas a magnitude is an ``arbitrary'' portion
of the SED. It is not fundamentally obvious that the luminosity within
a restricted wavelength range uniquely relates to the mass of a star.
Of course, these relations are well established and have been working very
well (not to mention that they only require photometry and no
spectroscopy).
After all, the infrared magnitudes are representative of the
luminosity of a star since similar relations between near-infrared magnitudes
and luminosity are also well established, and an ${\mathcal M}-L$ relation
is similarly fundamental as an ${\mathcal M}-R$ relation.
In any case, our work can be considered
an independent confirmation of the mass-magnitude relations of
\citet{2018arXiv181106938M}, \citet{2016AJ....152..141B},
or \citet{2000A&A...364..217D}.

It follows from the above discussion on the outliers that the age of a star
also plays a significant role in testing the validity of any method intended
for obtaining its mass and radius.
All of these,
except the PARSEC-based masses ${\mathcal M}_P$ (Section~\ref{sec:padova}), assume an age of
at least a few hundred megayears, i.e., that the stars have left the Hayashi track
and reached the main sequence. Our spectral analysis \citepalias[Section~\ref{sec:temperatures} and][]{pre013} 
assumes for the \logg\ determination an age of 5\,Gyr, which directly
enters the spectroscopic mass ${\mathcal M}_{\logg}$. Both the mass-radius based
masses ${\mathcal M}_{{\mathcal M}-R}$ and the photometric masses ${\mathcal M}_{{\mathcal M}-K_s}$
use empirical relations based on samples of old stars.
This is in turn an advantage of the PARSEC-based masses  ${\mathcal M}_P$
because, for example, the GTO targets of the $\beta$~Pictoris moving group and the Upper Scorpius OB association
have significantly lower ${\mathcal M}_P$ values than all the other masses.
At the same time the method for obtaining ${\mathcal M}_P$ yields
young ages for these sources.
However, a detailed investigation of the validity and consequences of the combined determination of
all parameters of M~dwarfs (and the CARMENES targets in particular) using this
method is beyond the scope of this paper and will be
discussed in a separate publication. Therefore, we do not list the ages determined with this method except for the
few examples above.
On the other hand, old stars such as the potential subdwarfs mentioned above
do not pose a problem for our method. In particular, as described
in Sec.~\ref{sec:mmr}, our mass-radius relation does not differ for different ages
above a few hundred million years, nor can we infer a metallicity
dependence down to metallicities of globular clusters.

Besides the age constraints inherent in the different methods, there are
further biases that enter some of the methods as already indicated in the
introduction. Both ${\mathcal M}_{{\mathcal M}-R}$
and  ${\mathcal M}_{{\mathcal M}-K_s}$ use empirical relations
based on binaries. However, the mass-radius relation for  ${\mathcal M}_{{\mathcal M}-R}$
uses close binaries, while the mass-luminosity relation for ${\mathcal M}_{{\mathcal M}-K_s}$ is
measured using wide binaries. The latter is certainly a better
representation for isolated M~dwarfs, i.e., our targets. If the
former suffers from, for example, inflated radii then there is a systematic
difference between ${\mathcal M}_{{\mathcal M}-R}$ and ${\mathcal M}_{{\mathcal M}-K_s}$.
Such inflated radii have been reported
(e.g., Jackson et al. 2018 or Kesseli et al. 2018, and for eclipsing binaries by Parsons et. al 2018),
\nocite{2018MNRAS.476.3245J,2018AJ....155..225K,2018MNRAS.481.1083P}
but only for young, magnetically active, or fast-rotating stars.
Our sample, however, mostly includes old, inactive, and slowly rotating stars (see Sec.~\ref{sec:sample}).
But if such a bias exists within the mass-radius relation derived in Sec.~\ref{sec:mmr},
it introduces an offset in our ${\mathcal M}_{{\mathcal M}-R}$ values.
As shown in Sec.~\ref{sec:kband} we find no offset between ${\mathcal M}_{{\mathcal M}-R}$ and ${\mathcal M}_{{\mathcal M}-K_s}$,
indicating no systematic difference between the mass-radius relation and
the mass-magnitude relation, or the samples on which they are based.

Finally, as described in Section~\ref{sec:temperatures}, the choice of a specific
set of synthetic spectra in our spectral analysis may introduce another systematic offset
on the order of our $\Delta\teff/\teff$ errors, or larger. But again, the agreement of
our ${\mathcal M}_{{\mathcal M}-R}$ with the ${\mathcal M}_{{\mathcal M}-K_s}$ that we observe indicates
that we did not introduce such a systematic offset.

\section{Conclusions}
\label{sec:conclusion}

\begin{figure}
\centering
\includegraphics[width=\hsize]{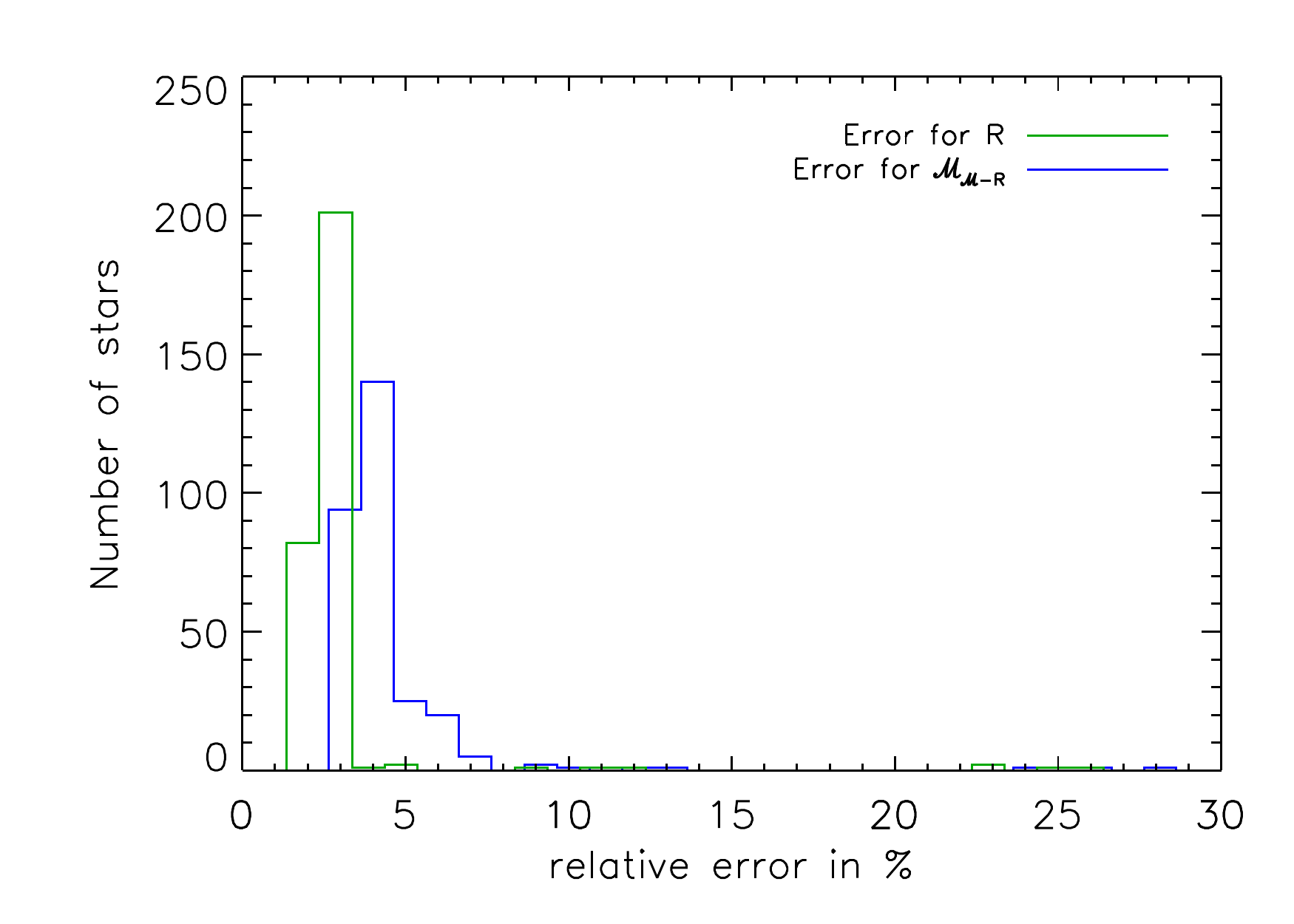}
   \caption{Histograms of the errors of $R$ and ${\mathcal M}_{{\mathcal M}-R}$.
           }
    \label{fig:errhist}
\end{figure}

We have derived radii and masses with individual error bars for 293 of the CARMENES
GTO M~dwarfs.
In particular, for measuring the masses, we used several methods and compared them.
For typical field stars that are not young, we list consistent masses with
all methods. For our main method we list radii $R$ that typically have errors of 2--3\%,
and masses ${\mathcal M}_{{\mathcal M}-R}$ that typically have errors of 3--5\% (cf.
Fig.~\ref{fig:errhist}).
Systematic uncertainties on the order of the error bars, however, cannot
be fully excluded.

For young stars,  ${\mathcal M}_{{\mathcal M}-R}$ and ${\mathcal M}_{{\mathcal M}-K_s}$
are not suitable since the applicable empirical relations are derived
for old stars (and cannot easily be adapted for young stars). Therefore, ${\mathcal M}_{{\mathcal M}-R}$
and ${\mathcal M}_{{\mathcal M}-K_s}$ for any young star in our sample (and not only for
the outliers discussed in Section~\ref{sec:outliers}) should be used with
care. The identification of all young stars, or determining their membership
in young moving groups, should it exist, is beyond the purpose of this paper and will
be published in the future. Similarly, an age determination of any star (not only those of our sample)
is challenging. For example, \citet{2018ApJ...863..166V} derived ages for some
of the CARMENES targets.
On the other hand,  ${\mathcal M}_P$ is not restricted to a single age,
and ages along with ${\mathcal M}_P$ can be determined simultaneously
as described in Section~\ref{sec:padova}.
Finally, when age estimates are available, our method of obtaining \logg\ will
be able to use isochrones for the appropriate age. Then our ${\mathcal M}_{\logg}$
is also an option, at least when we will be able to reduce the error bars
of \logg\ to be smaller than the current 0.07\,dex,
and when systematic effects of theoretical isochrones can be reduced.

Our methods for measuring $R$ and ${\mathcal M}_{{\mathcal M}-R}$ work best for a field
star of at least a few hundred million years when we can spectroscopically determine a reliable
\teff, when the parallax of the object is based on {\em Gaia} DR2,
and when $L$ is well anchored by {\em Gaia} and near-infrared photometry.
Then, we can determine for the target with the latest spectral type in our sample,
the M7\,V object Teegarden's star
$R=0.107\pm0.004\,R\subsun$ and ${\mathcal M}_{{\mathcal M}-R}=0.089\pm0.009\,{\mathcal M}\subsun$,
while we determine for our M0.0\,V targets average values
$R=0.566\pm0.016\,R\subsun$ and ${\mathcal M}_{{\mathcal M}-R}=0.574\pm0.021\,{\mathcal M}\subsun$.

\begin{acknowledgements}
We thank the anonymous referee for a very quick and very constructive report.
Furthermore, we thank T. Boyajian and A. W. Mann for helpful discussions during the preparation of this manuscript.
CARMENES is an instrument for the Centro Astron\'omico Hispano-Alem\'an de
  Calar Alto (CAHA, Almer\'{\i}a, Spain). 
  CARMENES is funded by the German Max-Planck-Gesellschaft (MPG), 
  the Spanish Consejo Superior de Investigaciones Cient\'{\i}ficas (CSIC),
  the European Union through FEDER/ERDF FICTS-2011-02 funds, 
  and the members of the CARMENES Consortium 
  (Max-Planck-Institut f\"ur Astronomie,
  Instituto de Astrof\'{\i}sica de Andaluc\'{\i}a,
  Landessternwarte K\"onigstuhl,
  Institut de Ci\`encies de l'Espai,
  Institut f\"ur Astrophysik G\"ottingen,
  Universidad Complutense de Madrid,
  Th\"uringer Landessternwarte Tautenburg,
  Instituto de Astrof\'{\i}sica de Canarias,
  Hamburger Sternwarte,
  Centro de Astrobiolog\'{\i}a and
  Centro Astron\'omico Hispano-Alem\'an), 
  with additional contributions by the Spanish Ministry of Science through projects AYA2016-79425-C3-1/2/3-P, ESP2016-80435-C2-1-R, AYA2015-69350-C3-2-P, and AYA2018-84089, 
  the Spanish  Ministerio de Educaci{\'o}n y Formaci{\'o}n Profesional through fellowship FPU15/01476,
  the German Science Foundation through the Major Research Instrumentation 
    Programme and DFG Research Unit FOR2544 ``Blue Planets around Red Stars'', 
  the Klaus Tschira Stiftung, 
  the states of Baden-W\"urttemberg and Niedersachsen, 
  and by the Junta de Andaluc\'{\i}a.
  CdB acknowledges the funding of his sabbatical position through the
   Mexican national council for science and technology (CONACYT grant CVU
   No. 448248).
     This publication makes use of VOSA, developed under the Spanish
     Virtual Observatory project,      the VizieR catalog access tool
     \citep{2000A&AS..143...23O} and the SIMBAD database \citep{2000A&AS..143....9W},
     operated both at CDS, Strasbourg, France, and the interactive
     graphical viewer and editor for tabular data TOPCAT \citep{2005ASPC..347...29T}.

\end{acknowledgements}

\bibliographystyle{aa}

\clearpage
\begin{appendix}

\section{Additional plots}

\begin{figure} \centering
\includegraphics[width=\hsize]{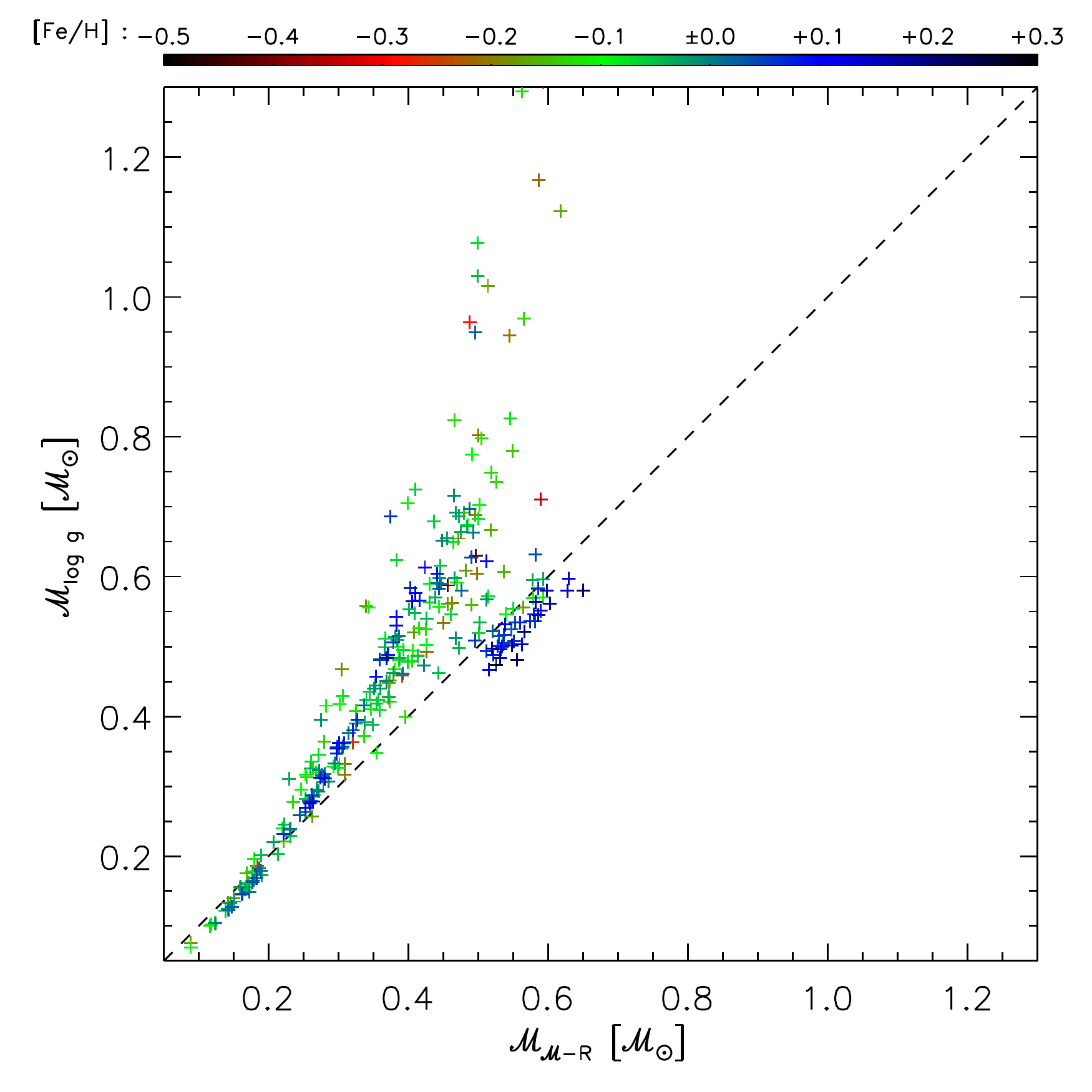}
   \caption{Same as Fig.~\ref{fig:mmr_mlogg}, but without error bars for better clarity
           }
    \label{fig:mmr_mlogg_wo}
\end{figure}

In this appendix, we provide supplemental material.
First, we repeat Fig.~\ref{fig:mmr_mlogg} but we use linear axes and omit
the error bars in Fig.~\ref{fig:mmr_mlogg_wo} for clarity.

\begin{figure}
\centering
\includegraphics[width=\hsize]{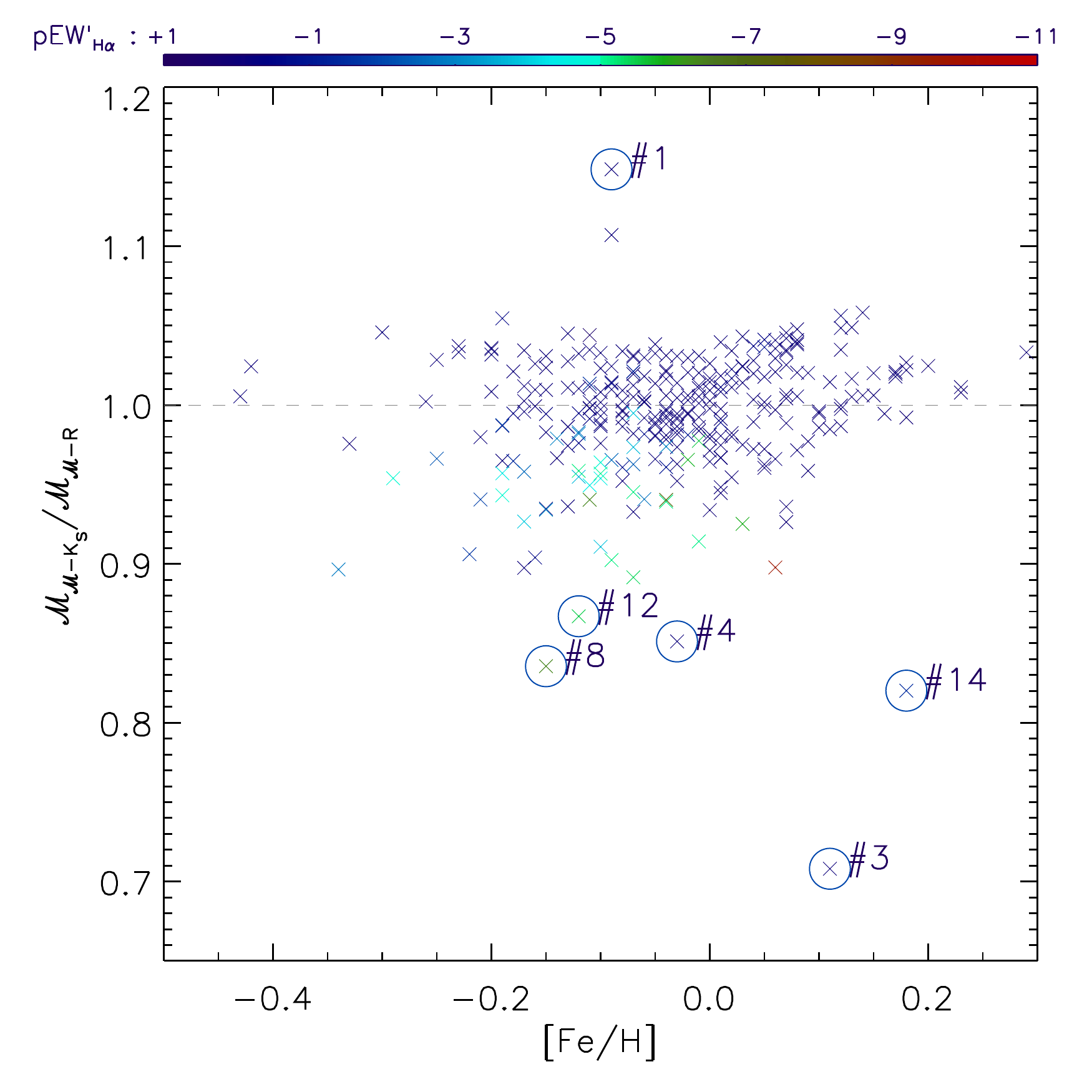}
   \caption{Ratio of the masses of Fig.~\ref{fig:mmr} as a function of [Fe/H]
             using the pseudo equivalent width pEW$^\prime_{{\rm H}\alpha}$
            of \citet{pre034} for the color coding. Negative values of pEW$^\prime_{{\rm H}\alpha}$ denote H$\alpha$ emission.
            Error bars are omitted for clarity.
            See Fig.~\ref{fig:mmr} for the identification of the encircled stars.
            }
    \label{fig:mmr_rel}
\end{figure}

Next, we resolve the confusion seen in 
Fig.~\ref{fig:mmr} where all stars lie very close to the 1-1-line when
plotting ${\mathcal M}_{{\mathcal M}-K_s}$ against ${\mathcal M}_{{\mathcal M}-R}$.
Instead, in Fig.~\ref{fig:mmr_rel} we now plot the ratio ${\mathcal M}_{{\mathcal M}-K_s}$ over ${\mathcal M}_{{\mathcal M}-R}$
against the metallicity from Section~\ref{sec:temperatures} and also take into
account the activity of the stars using the pseudo equivalent width pEW$^\prime_{{\rm H}\alpha}$
of \citet{pre034} for the color coding.
We can now identify that all
H$\alpha$ emittiers
have slightly higher ${\mathcal M}_{{\mathcal M}-R}$ masses (and by Eqs.~\ref{eq:mmr} and \ref{eq:mrm}
also slightly larger radii) compared to the ${\mathcal M}_{{\mathcal M}-K_s}$ masses.
The H$\alpha$ absorbers, however, are symmetrically spread around a ratio of unity.
There is a trend that the ${\mathcal M}_{{\mathcal M}-R}$ masses
are lower than the ${\mathcal M}_{{\mathcal M}-K_s}$ masses only for the highest metallicities. A similar comparison of ${\mathcal M}_{{\mathcal M}-R}$
with the other alternative masses ${\mathcal M}_{\logg}$ and ${\mathcal M}_P$, however,
does neither show a trend with activity nor with metallicity in their deviation from ${\mathcal M}_{{\mathcal M}-R}$.

\begin{figure}
\centering
\includegraphics[width=\hsize]{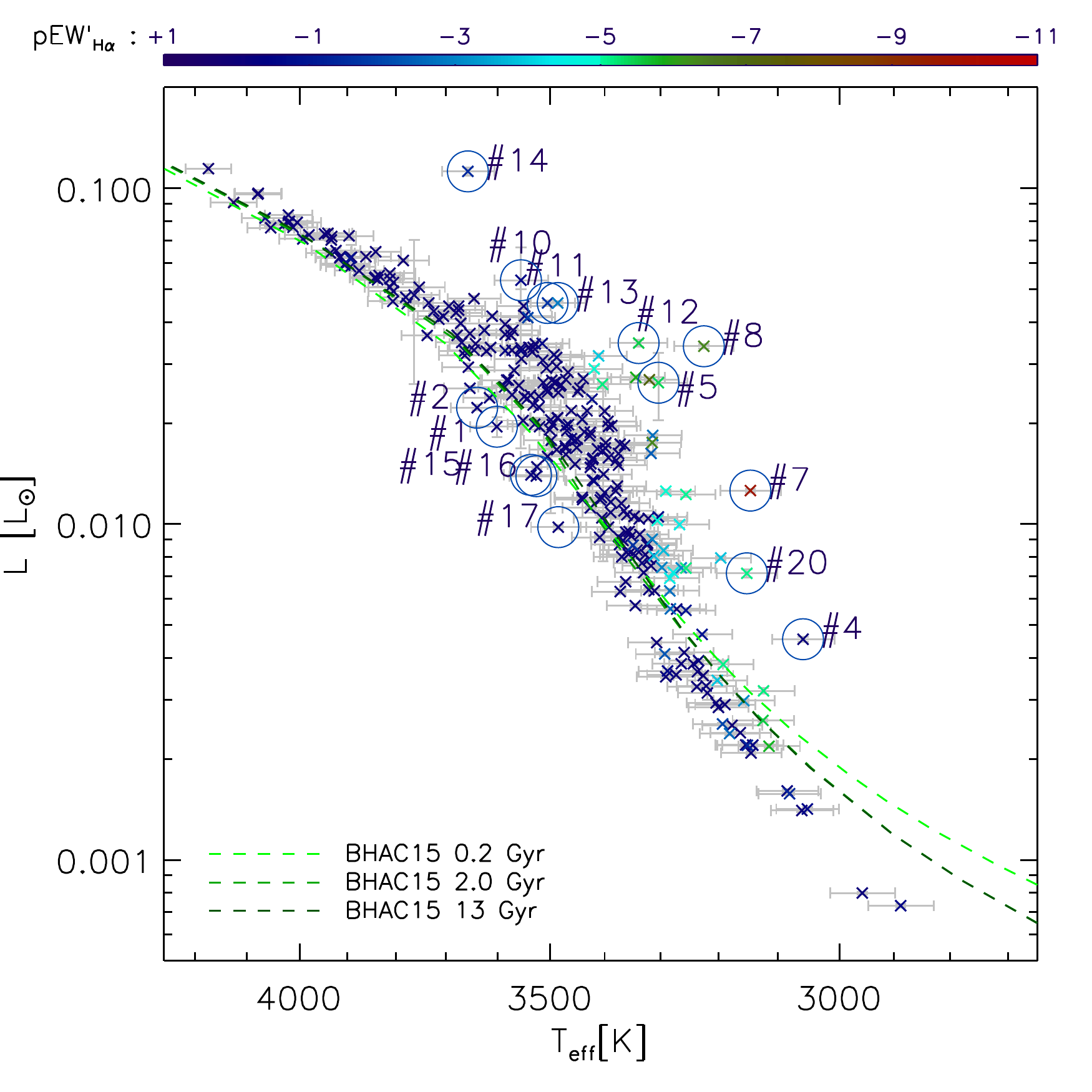}
   \caption{Same as Fig.~\ref{fig:hrd}, but using the pseudo equivalent width pEW$^\prime_{{\rm H}\alpha}$
            of \citet{pre034} for the color coding. Negative values of pEW$^\prime_{{\rm H}\alpha}$ denote H$\alpha$ emission.
            See Fig.~\ref{fig:hrd} for the identification of the encircled stars.
            }
      \label{fig:hrdhalpha}
\end{figure}

\begin{figure}
\centering
\includegraphics[width=\hsize]{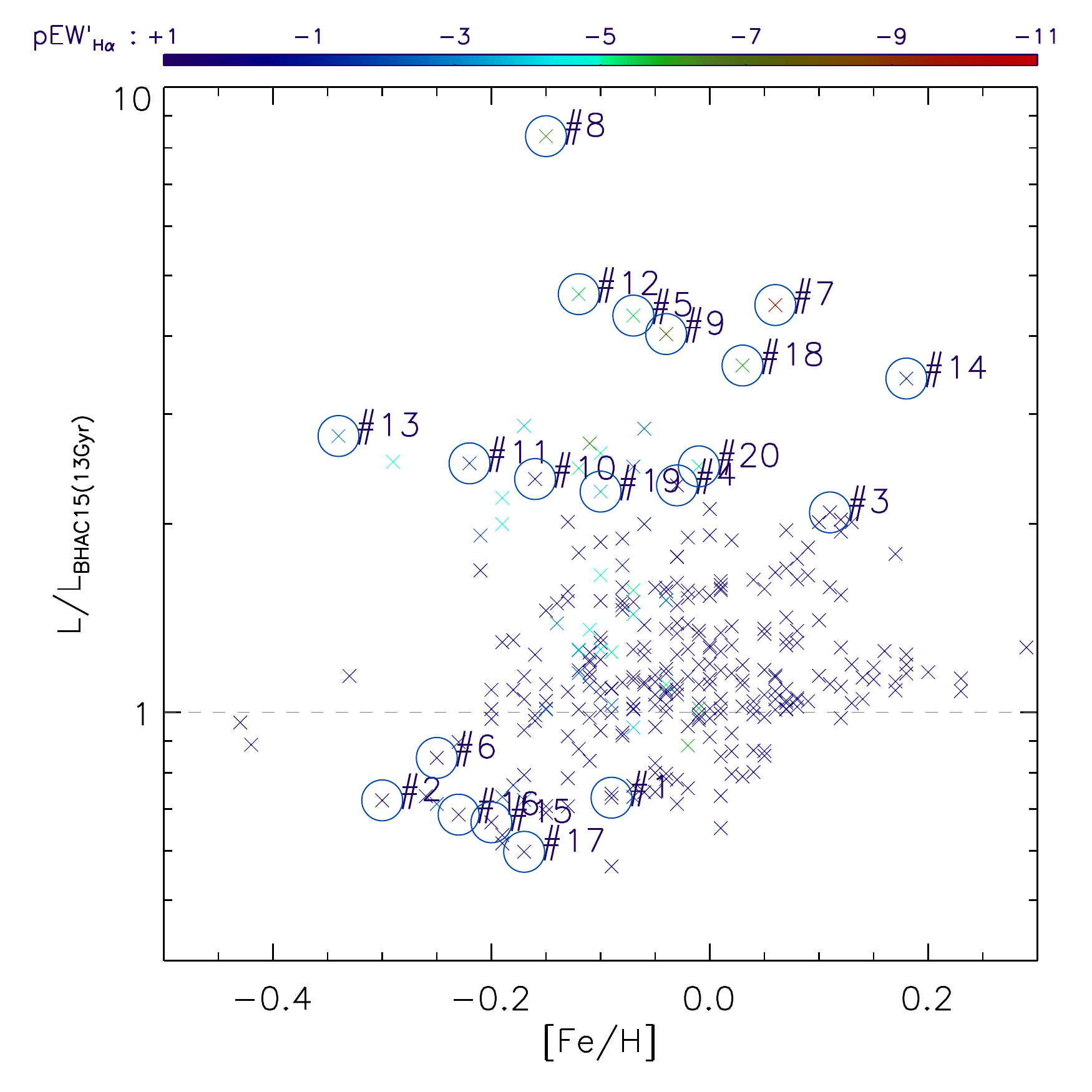}
   \caption{Ratio of the luminosity of the HRD in Fig.~\ref{fig:hrd} and the luminosity of a 13\,Gyr
                isochrone from \citetalias{2015A&A...577A..42B} for the measured effective temperatures as a function of [Fe/H]
             using the pseudo equivalent width pEW$^\prime_{{\rm H}\alpha}$
            of \citet{pre034} for the color coding. Negative values of pEW$^\prime_{{\rm H}\alpha}$ denote H$\alpha$ emission.
            Error bars are omitted for clarity.
            See Fig.~\ref{fig:hrd} for the identification of the encircled stars.
            }
    \label{fig:hrd_rel}
\end{figure}

We also repeat the HRD of Fig.~\ref{fig:hrd} but using the pseudo equivalent width pEW$^\prime_{{\rm H}\alpha}$
of \citet{pre034} for the color coding in Fig.~\ref{fig:hrdhalpha}.
We again magnify the spread seen in the HRDs in Figs.~\ref{fig:hrd} and \ref{fig:hrdhalpha}
by plotting the ratio $L$ over $L_{\rm BHAC15(13\,Gyr)}$ against
the metallicity from Section~\ref{sec:temperatures} in Fig.~\ref{fig:hrd_rel} and again taking into
account the activity of the stars using the pseudo equivalent width pEW$^\prime_{{\rm H}\alpha}$
of \citet{pre034} for the color coding. This $L_{\rm BHAC15(13\,Gyr)}$ is the luminosity of a
star on the 13\,Gyr isochrone from \citetalias{2015A&A...577A..42B} selected by its
measured \teff, i.e., the x-axes of the HRD in Fig.~\ref{fig:hrd}.
Similar to Figs.~\ref{fig:hrd} and \ref{fig:hrdhalpha} we can identify that
the active stars have the largest discrepancy from a theoretical, 13\,Gyr old $L-\teff$ sequence,
and we can also see that all stars with [Fe/H]$\gtrsim$0.5\,dex are above
the ratio of unity. The same behavior could be seen in a plot
that magnifies the spread seen in Figs.~\ref{fig:mteff} and \ref{fig:mteffhalpha}
in an equivalent fashion but since no new information is revealed we omit such a plot.

\begin{figure} \centering
\includegraphics[width=\hsize]{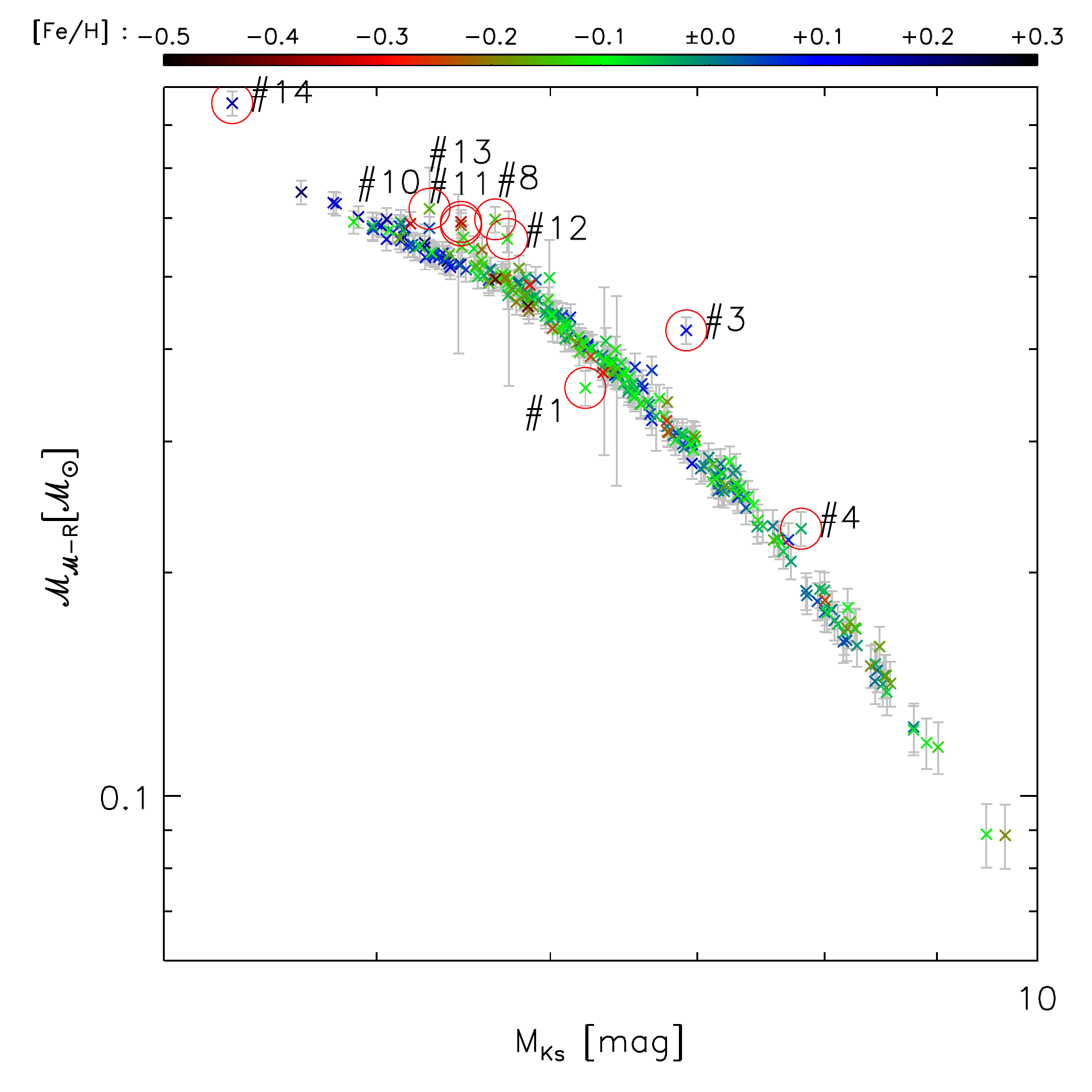}
   \caption{ Value of ${\mathcal M}_{{\mathcal M}-R}$ against absolute $K_s$ magnitude $M_{K_s}$.
          \protect Obvious outliers, and stars discussed in reference to this figure in Section~\ref{sec:outliers}, are encircled:
\#1=Gl~411~(J11033+359), \#3=GJ~4063~(J18346+401), \#4=GJ~1235~(J19216+208), \#8=1RXS~J050156.7+010845~(J05019+011), \#10=G~234-057~(J09133+688), \#11=TYC~3529-1437-1~(J18174+483), \#12=StKM~2-809~(J12156+526), \#13=GJ~9520~(J15218+209), and \#14=K2-33~(J16102-193).
             Outliers \#11 and \#13 lie almost on top of each other.
            }
    \label{fig:mks}
\end{figure}

\begin{figure} \centering
\includegraphics[width=\hsize]{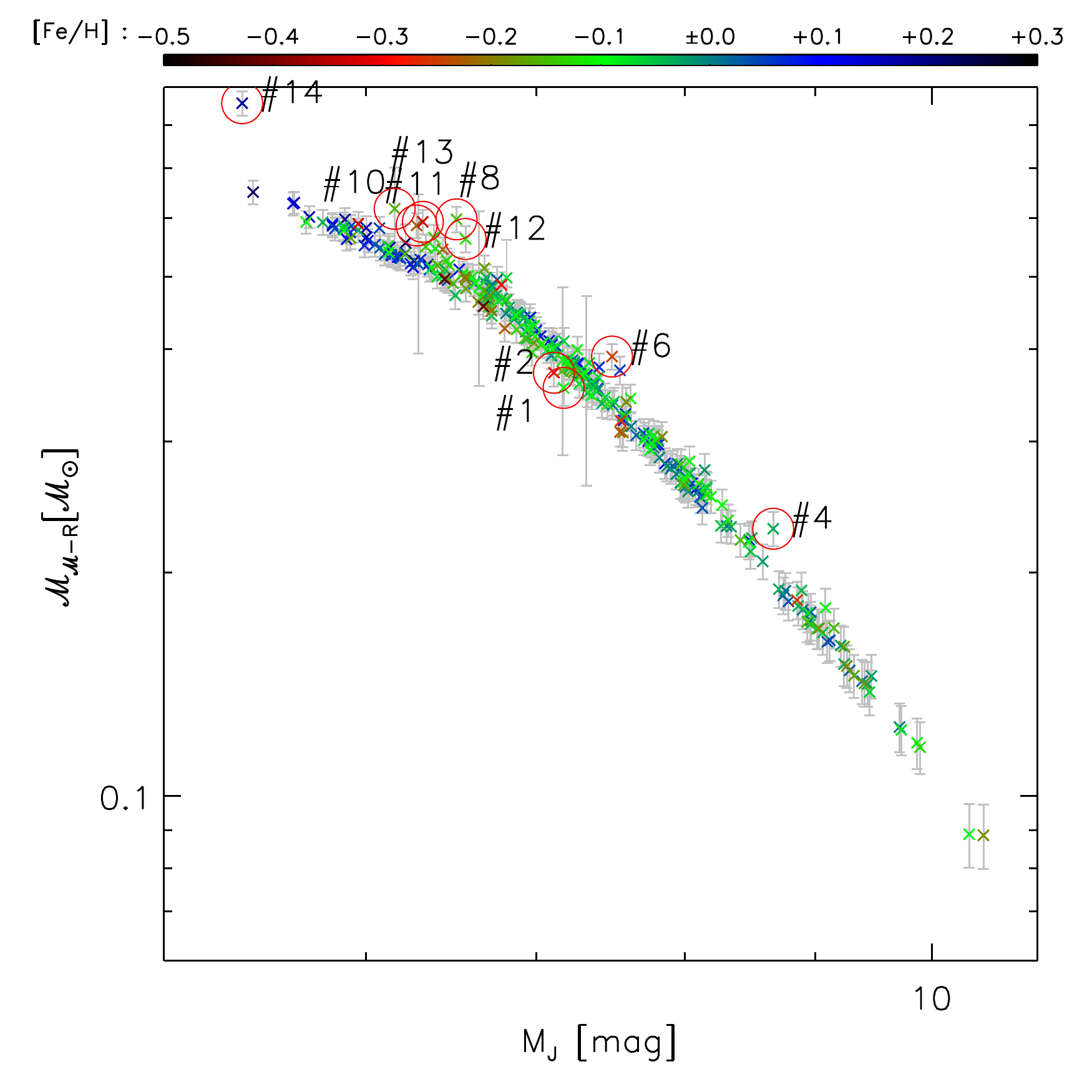}
   \caption{Value of ${\mathcal M}_{{\mathcal M}-R}$ against absolute $J$ magnitude $M_J$.
          \protect Obvious outliers, and stars discussed in reference to this figure
in Section~\ref{sec:outliers}, are encircled:
\#1=Gl~411~(J11033+359), \#2=Gl~412A~(J11054+435), \#4=GJ~1235~(J19216+208), \#6=Gl~15A~(J00183+440), \#8=1RXS~J050156.7+010845~(J05019+011), \#10=G~234-057~(J09133+688), \#11=TYC~3529-1437-1~(J18174+483), \#12=StKM~2-809~(J12156+526), \#13=GJ~9520~(J15218+209), and \#14=K2-33~(J16102-193).
             Outliers \#11 and \#13 lie almost on top of each other.
              }
    \label{fig:mj}
\end{figure}

We finally plot ${\mathcal M}_{{\mathcal M}-R}$
against absolute magnitudes $K_s$ and $J$ in Figs.~\ref{fig:mks} and \ref{fig:mj}.
These two plots help in identifying the outliers of Section~\ref{sec:outliers}.

\section{Tables}

\begin{table*}
\longtab[1]{
\small
\begin{longtable}{l  r@{$\pm$}l r@{$\pm$}l r@{$\pm$}l r@{$\pm$}l r@{$\pm$}l r@{$\pm$}l r@{$\pm$}l}
\caption{\label{tab:results} Stellar parameters\tablefootmark{a} for all stars of our sample.}\\
\hline\hline
  \multicolumn{1}{l}{Karmn} &
  \multicolumn{2}{c}{\teff\ [K]} & 
  \multicolumn{2}{c}{ \logg} & 
  \multicolumn{2}{c}{\zfe} &
  \multicolumn{2}{c}{$L$ [$L\subsun$]} &
  \multicolumn{2}{c}{$R$ [$R\subsun$]} & 
  \multicolumn{2}{c}{${\mathcal M}_{{\mathcal M}-R}$ [${\mathcal M}\subsun$]}  &
  \multicolumn{2}{c}{${\mathcal M}_{{\mathcal M}-K_s}$ [${\mathcal M}\subsun$]} \\
  \multicolumn{1}{l}{\quad Name} &
  \multicolumn{2}{c}{SpT} & 
  \multicolumn{2}{c}{$\logg_c$} & 
  \multicolumn{2}{c}{} &
  \multicolumn{2}{c}{$d$ [pc]} &
  \multicolumn{2}{c}{$R_{P}$ [$R\subsun$]}  &
  \multicolumn{2}{c}{${\mathcal M}_{\logg}$ [${\mathcal M}\subsun$]} &
  \multicolumn{2}{c}{${\mathcal M}_{P}$ [${\mathcal M}\subsun$]} \\
\hline
\endfirsthead
\caption{Stellar parameters\tablefootmark{a} for all stars of our sample (continued).}\\
\hline\hline
  \multicolumn{1}{l}{Karmn} &
  \multicolumn{2}{c}{\teff\ [K]} & 
  \multicolumn{2}{c}{ \logg} & 
  \multicolumn{2}{c}{\zfe} &
  \multicolumn{2}{c}{$L$ [$L\subsun$]} &
  \multicolumn{2}{c}{$R$ [$R\subsun$]} & 
  \multicolumn{2}{c}{${\mathcal M}_{{\mathcal M}-R}$ [${\mathcal M}\subsun$]}  &
  \multicolumn{2}{c}{${\mathcal M}_{{\mathcal M}-K_s}$ [${\mathcal M}\subsun$]} \\
  \multicolumn{1}{l}{\quad Name} &
  \multicolumn{2}{c}{SpT} & 
  \multicolumn{2}{c}{$\logg_c$} & 
  \multicolumn{2}{c}{} &
  \multicolumn{2}{c}{$d$ [pc]} &
  \multicolumn{2}{c}{$R_{P}$ [$R\subsun$]}  &
  \multicolumn{2}{c}{${\mathcal M}_{\logg}$ [${\mathcal M}\subsun$]} &
  \multicolumn{2}{c}{${\mathcal M}_{P}$ [${\mathcal M}\subsun$]} \\
\hline
\endhead
\hline
\endfoot 
\makeatletter{}J00051+457 & 3675 & 51 & 4.84 & 0.07 & -0.16 & 0.16 & 0.04341 & 0.00086 & 0.514 & 0.015 & 0.518 & 0.020 & 0.518 & 0.013\\
\quad Gl 2 & \multicolumn{2}{c}{dM1.0\,V} & 4.731 & 0.030 & \multicolumn{2}{c}{ }  &  11.500 &  0.005\tablefootmark{}   & \multicolumn{2}{c}{\dots}  & 0.667 & 0.114 & \multicolumn{2}{c}{\dots} \\[0.5ex]
J00067-075 & 3061 & 51 & 5.19 & 0.07 & -0.13 & 0.16 & 0.00140 & 0.00002 & 0.133 & 0.005 & 0.116 & 0.009 & 0.115 & 0.003\\
\quad GJ 1002 & \multicolumn{2}{c}{dM5.5\,V} & 5.255 & 0.045 & \multicolumn{2}{c}{ }  &   4.849 &  0.003\tablefootmark{}   & 0.152 & 0.002 & 0.100 & 0.018 & 0.123 & 0.002\\[0.5ex]
J00162+198E & 3323 & 51 & 5.03 & 0.07 &  0.08 & 0.16 & 0.00877 & 0.00013 & 0.282 & 0.009 & 0.274 & 0.013 & 0.266 & 0.007\\
\quad GJ 1006B & \multicolumn{2}{c}{dM4.0\,V} & 4.973 & 0.034 & \multicolumn{2}{c}{ }  &  15.326 &  0.019\tablefootmark{}   & 0.318 & 0.004 & 0.312 & 0.054 & 0.310 & 0.006\\[0.5ex]
J00183+440 & 3615 & 51 & 4.91 & 0.07 & -0.25 & 0.16 & 0.02379 & 0.00026 & 0.393 & 0.011 & 0.391 & 0.016\tablefootmark{Out} & 0.402 & 0.010\\
\quad Gl 15A & \multicolumn{2}{c}{dM1.0\,V} & 4.841 & 0.030 & \multicolumn{2}{c}{ }  &   3.563 &  0.001\tablefootmark{}   & 0.363 & 0.038 & 0.458 & 0.078 & 0.369 & 0.041\\
\end{longtable}
{\bf Notes. }
{Also available in electronic form at the CDS via anonymous ftp to cdsarc.u-strasbg.fr (130.79.128.5)
or via \url{http://cdsweb.u-strasbg.fr/cgi-bin/qcat?J/A+A/}}\\
\tablefoottext{a}{\teff, \logg, $\logg_c$, \zfe, $L$, $R$, $R_{P}$, ${\mathcal M}_{{\mathcal M}-R}$, ${\mathcal M}_{\logg}$,
${\mathcal M}_{{\mathcal M}-K_s}$ and ${\mathcal M}_{P}$ are determined in this work.}
{Spectral types are from \citet{2015A&A...577A.128A} and references therein.}
{Heliocentric distances are from \citet{2018arXiv180409365G} except
                  where indicated otherwise}:
\tablefoottext{HIP}{\citet{2007A&A...474..653V}};
\tablefoottext{Yale}{\citet{1995gcts.book.....V}};
\tablefoottext{URAT}{\citet{2016AJ....151..160F}};
\tablefoottext{CC17}{Following \citet{2017A&A...597A..47C}.}\\
\tablefoottext{New}{New stars not tabulated by \citet{pre021}.}\\
\tablefoottext{Out}{Masses of outliers discussed in Section~\ref{sec:outliers}, to be used with care.}
}

\end{table*}

\longtab[2]{
\begin{table*}
\small
\caption{Masses and radii of the eclipsing M~dwarf binaries used to derive  Eq.~\ref{eq:mmr}.}
\label{tab:mr}
\centering
\begin{tabular}{l r r l r@{$\pm$}l r@{$\pm$}l r}
\hline\hline
Name     &  RA (J2000) & Dec (J2000) & SpT\tablefootmark{a} & \multicolumn{2}{c}{${\mathcal M}$                        }  &  \multicolumn{2}{c}{$R$             } & \quad Ref. \\
         &             &             &                      & \multicolumn{2}{c}{               [${\mathcal M}\subsun$]}  &  \multicolumn{2}{c}{    [$R\subsun$]} &            \\
\hline
OGLE-TR-122\,B 		&  11 06 51.89 & -60 51 45.9 &	M &	0.092 &	0.009 	&	0.120 &	0.018 & 16 \\
WTS 19g-4-02069\,B	&  19 35 03.55 & +36 31 16.5 &  M     & 0.143 &  0.006   &       0.174 &  0.006 & 20  \\
LP 837-20\,B   		&  05 44 57.93 & -24 56 09.7 &  M     & 0.179 &  0.002   &       0.218 &  0.011 & 19 \\
LP 661-13\,B   		&  06 56 18.95 & -08 35 46.5 &	M&	0.19400& 0.00034 &       0.2174 & 0.0023& 18 \\
Kepler-16\,B 		&  19 16 18.18 & +51 45 26.8 &	M &	0.1959&	0.0031 	&	0.2262& 0.00056& 4 \\
OGLE-TR-125\,B 		&  10 57 51.86 & -61 43 58.9 &	M &	0.209 &	0.033 	&	0.211 &	0.027 & 16 \\
KOI-126\,C 		&  19 49 54.20 & +41 06 51.4 &	M &	0.2127&	0.0026 	&	0.2318 &0.0013 &7 \\
CM Dra\,B 		&  16 34 20.33 & +57 09 44.4 &	M4.5 & 	0.2141&	0.0010 	&	0.2396& 0.0015& 10 \\
CM Dra\,A 		&  16 34 20.33 & +57 09 44.4 &	M4.5 & 	0.2310&	0.0009 	&	0.2534& 0.0019& 10 \\
SDSS-MEB-1\,B 		&  03 18 23.88 & -01 00 18.4 &	M &	0.240 &	0.022 	&	0.248 &	0.009 &	5 \\
KOI-126\,B 		&  19 49 54.20 & +41 06 51.4 &	M &	0.2413&	0.003 	&	0.2543 &0.0014 &7 \\
LP 837-20\,A     		&  05 44 57.93 & -24 56 09.7 &  M     & 0.244 &  0.003   &       0.261  & 0.009 & 19 \\
1RXS J154727.5+450803\,A	&  15 47 27.41 & +45 07 51.3 &	M4&	0.2576&	0.0085 	&	0.2895 &0.0068 &8 \\
1RXS J154727.5+450803\,B	&  15 47 27.41 & +45 07 51.3 &	M4&	0.2585&	0.0080 	&	0.2895 &0.0068 &8 \\
OGLE-TR-5\,B 		&  17 51 49.34 & -30 01 44.4 &	M &	0.271& 	0.035 	&	0.263 &	0.012 & 15 \\
SDSS-MEB-1\,A	 	&  03 18 23.88 & -01 00 18.4 &	M &	0.272 &	0.020 	&	0.268 &	0.010 &	5 \\
HAT-TR-318-007\,B 	&  08 50 32.96 & +12 08 23.6 &  M5    & 0.2721&  0.0042  &       0.2913 & 0.0024& 24 \\
LSPM J1112+7626\,B	&  11 12 42.34 & +76 26 56.4 &	M &	0.2745 &0.0012 	&	0.2978 &0.0050 &1 \\
OGLE-TR-7\,B 		&  17 52 08.66 & -29 56 12.1 &	M &	0.281 &	0.029 	&	0.282 &	0.013 & 15 \\
LP 661-13\,A   		&  06 56 18.95 & -08 35 46.5 &	M3.5&	0.30795& 0.00084 &       0.3226 & 0.0033& 18     \\
LP 133-373\,A 		&  14 04 08.89 & +50 20 38.7 &	M4& 	0.340 &	0.014 	&	0.33 &	0.02 &	6 \\
LP 133-373\,B 		&  14 04 08.89 & +50 20 38.7 &	M4& 	0.340 &	0.014 	&	0.33 &	0.02 &	6 \\
OGLE-TR-6\,B 		&  17 51 03.07 & -29 55 49.8 &	M &	0.359 &	0.025 	&	0.393 &	0.018 & 15 \\
T-Cyg1-12664\,B  		&  19 51 39.82 & +48 19 55.4 &  M&      0.376 &  0.017   &       0.3475 & 0.0081& 17 \\
MOTESS-GNAT 2056316\,B 	&  23 14 38.16 & +03 39 49.4 &	M3&   	0.382 &	0.0023 	&	0.374 &	0.0035 &2 \\
OGLE-TR-18\,B 		&  17 54 16.46 & -29 43 11.9 &	M &	0.387 &	0.049 	&	0.390 &	0.040 & 15 \\
LSPM J1112+7626\,A	&  11 12 42.34 & +76 26 56.4 &	M4   &	0.3946 &0.0023 	&	0.3860 &0.0050 &1 \\
CU Cnc\,B 		&  08 31 37.57 & +19 23 39.4 &	M3.5 & 	0.399 &	0.0015 	&	0.391 &	0.0104& 12 \\
CU Cnc\,A 		&  08 31 37.57 & +19 23 39.4 &	M3.5 & 	0.433 &	0.0017 	&	0.432 &	0.0052& 12 \\
MOTESS-GNAT 646680\,B 	&  10 30 55.21 & +03 34 26.7 &	M2&   	0.443 &	0.0020 	&	0.427 &	0.0061 &2 \\
HAT-TR-318-007\,A 	&  08 50 32.96 & +12 08 23.6 &  M4    & 0.448 &  0.001   &       0.4548 & 0.0036& 24 \\
MOTESS-GNAT 2056316\,A 	&  23 14 38.16 & +03 39 49.4 &	M2&   	0.469 &	0.0023 	&	0.441 &	0.0035 &2 \\
Tres-Her0-07621\,B	&  16 50 20.73 & +46 39 01.4 & 	M &	0.489 &	0.003 	&	0.452 &	0.06  & 13 \\
MOTESS-GNAT 78457\,B 	&  03 26 20.73 & +03 12 36.3 &	M4&   	0.491 &	0.0018 	&	0.471 &	0.0111 &2 \\
Tres-Her0-07621\,A	&  16 50 20.73 & +46 39 01.4 & 	M &	0.493 &	0.003 	&	0.453 &	0.06  & 13 \\
NSVS 01031772\,B 	&  13 45 34.87 & +79 23 48.3 &	M & 	0.498 &	0.0022 	&	0.509 &	0.0026 &25 \\
MOTESS-GNAT 646680\,A 	&  10 30 55.21 & +03 34 26.7 &	M1&   	0.499 &	0.0020 	&	0.457 &	0.0070 &2 \\
MOTESS-GNAT 78457\,A 	&  03 26 20.73 & +03 12 36.3 &	M3&   	0.527 &	0.0018 	&	0.505 &	0.0111 &2 \\
WTS 19g-4-02069\,A	&  19 35 03.55 & +36 31 16.5 &  M3.5  & 0.530 &  0.020   &       0.510 &  0.010 & 20  \\
MOTESS-GNAT 116309\,B 	&  04 48 09.63 & +03 17 48.1 &	M0&   	0.531 &	0.0016 	&	0.532 &	0.0073 &2 \\
MOTESS-GNAT 506664\,B 	&  07 43 11.57 & +03 16 22.1 &	M2&   	0.544 &	0.0016 	&	0.513 &	0.0068 &2 \\
MOTESS-GNAT 116309\,A 	&  04 48 09.63 & +03 17 48.1 &	K8&   	0.567 &	0.0015 	&	0.552 &	0.0102 &2 \\
MOTESS-GNAT 506664\,A 	&  07 43 11.57 & +03 16 22.1 &	M1&   	0.584 &	0.0015 	&	0.560 &	0.0039 &2 \\
V530 Ori\,B		&  06 04 33.8\,\,\, & -03 11 52\,\,\,\, &       M1    &	0.5955&  0.0022	&	0.5873 & 0.0067& 22 \\
YY Gem\,A 		&  07 34 37.58 & +31 52 11.1 &	M1   &	0.597 &	0.0034 	&	0.619 &	0.0040& 11 \\
GU Boo\,B 		&  15 21 54.82 & +33 56 09.1 &	M1&   	0.599 &	0.0044 	&	0.620 &	0.014 &	9 \\
YY Gem\,B 		&  07 34 37.58 & +31 52 11.1 &	M1   &	0.601 &	0.0034 	&	0.603&	0.0041& 11 \\
GU Boo\,A 		&  15 21 54.82 & +33 56 09.1 &	M1&   	0.609 &	0.0050 	&	0.623 &	0.0112& 9 \\
2MASSJ01542930+0053266\,B &  01 54 29.30 & +00 53 26.7 &	M1& 	0.62 &	0.03 	&	0.61 &	0.09  & 14 \\
Kepler-16\,A 		&  19 16 18.18 & +51 45 26.8 &	K7 &	0.654 &	0.017 	&	0.6489& 0.0013 &4 \\
2MASSJ01542930+0053266\,A &  01 54 29.30 & +00 53 26.7 &	M0& 	0.66 &	0.03 	&	0.64 &	0.08  & 14 \\
KIC 6131659\,B    	&  19 37 06.97 & +41 26 12.8 &  M     & 0.685 &  0.005   &       0.6395&  0.0061& 21 \\
ASAS J082552-1622.8\,B	&  08 25 51.61 & -16 22 47.4 &  M0    & 0.6872&  0.0049	&	0.699 &  0.013 & 23 \\
RXJ0239.1-1028\,B	&  02 39 08.75 & -10 27 46.4 &	K7 &	0.693 &	0.006 	&	0.703 &	0.002 &	3 \\
RXJ0239.1-1028\,A	&  02 39 08.75 & -10 27 46.4 &	K7 &	0.730 &	0.009 	&	0.741 &	0.004 &	3 \\
 \hline
\end{tabular}

\flushleft
{\bf Notes. }
\tablefoottext{a}{The spectral types are the ones listed in SIMBAD or as given in the corresponding
reference otherwise.}
\tablebib{
(1)~\citet{2011ApJ...742..123I};
(2)~\citet{2011ApJ...728...48K}; (3)~\citet{2007ASPC..362...26L}; (4)~\citet{2012ApJ...751L..31B}; (5)~\citet{2008ApJ...684..635B}; (6)~\citet{2007ApJ...661.1112V}; (7)~\citet{2011Sci...331..562C}; (8)~\citet{2011AJ....141..166H}; (9)~\citet{2005ApJ...631.1120L}; (10)~\citet{2009ApJ...691.1400M}; (11)~\citet{2002ApJ...567.1140T}; (12)~\citet{2003A&A...398..239R}; (13)~\citet{2005ApJ...625L.127C}; (14)~\citet{2008MNRAS.386..416B}; (15)~\citet{2005A&A...431.1105B}; (16)~\citet{2005A&A...438.1123P}; (17)~\citet{2017A&A...600A..55I}; (18)~\citet{2017ApJ...836..124D}; (19)~\citet{2015MNRAS.451.2263Z}; (20)~\citet{2013MNRAS.431.3240N}; (21)~\citet{2012ApJ...761..157B}; (22)~\citet{2014ApJ...797...31T}; (23)~\citet{2011A&A...526A..29H}; (24)~\citet{2018arXiv180103570H}; (25)~\citet{2006astro.ph.10225L}. }
\end{table*}
}

\end{appendix}

\end{document}